\documentclass[12pt,a4paper]{article}
\usepackage{graphicx}
\usepackage{bm}
\def\ben{\begin{enumerate}}  \def\een{\end{enumerate}}
\def\beq{\begin{equation}}   \def\eeq{\end{equation}}
\def\gbar{\relax\ifmmode\overline{g}\else{$\overline{g}${ }}\fi}
\def\albar{\overline \alpha}       \def\etc{\hbox{\it etc.}{ }}
\def\define{\buildrel \rm def \over =}
\begin{document}
\begin{flushright}
 {\small Contribution to the proceedings of conference ``RG 2000" \\
    (Taxco, Mexico, Jan. 1999). To be published in {\sf Phys. Repts};\\
     JINR preprint E2-2000-9}
\end{flushright}
 \vspace{5mm}

\begin{center}
{\large\bf BOGOLIUBOV Renormalization Group and \\ Symmetry of
Solution in Mathematical Physics }
\end{center}
 \vspace{2mm} %
\begin{flushright}
 {\textit{Dedicated to the memory of Boris Medvedev}}
\end{flushright}
 \vspace{2mm} %
\begin{center}
Dmitrij V. Shirkov  \\
\vspace{2mm} %
{\footnotesize\it Bogoliubov Laboratory, JINR, Dubna, 141980
Russia; \ e-mail: shirkovd@thsun1.jinr.ru } \\ \vspace{2mm} and \\
Vladimir F. Kovalev\\
\vspace{2mm}%
{\footnotesize\it Institute for Mathematical Modelling, Moscow,
125047 Russia; \ e-mail: kovalev@imamod.ru} \\
\end{center}
\vspace{2mm}
\begin{abstract}
Evolution of the concept known in the theoretical physics as the
Renormalization Group (RG) is presented. The corresponding
symmetry, that has been first introduced in QFT in mid-fifties, is
a continuous symmetry of a solution with respect to transformation
involving parameters (e.g., of boundary condition) specifying some
particular solution.

After short detour into Wilson's discrete semi-group, we follow
the expansion of QFT RG and argue that the underlying
transformation, being considered as a reparameterisation one, is
closely related to the self-similarity property. It can be treated
as its generalization, the {\sf Functional Self-similarity} (FS).

Then, we review the essential progress during the last decade of
the FS concept in application to {\sf boundary value problem
formulated in terms of differential equations}. A summary of a
regular approach recently devised for discovering the RG = FS
symmetries with the help of the modern Lie group analysis and some
of its applications are given.

  As a main physical illustration, we give application of new approach to
solution for a problem of {\sf self--focusing laser beam} in a
non-linear medium.
\end{abstract}
\newpage
\tableofcontents
\newpage

\section{The Bogoliubov renormalization group }

\subsection{Historical introduction \label{ss1.1}}

\subsubsection{\protect\small\it{The discovery of the renormalization
group}}\label{sss1.1.1}

 In 1952-1953 St\"uckelberg and Petermann \cite{stp} discovered\footnote{
For a more detailed exposition of the RG early history, see our
recent reviews \cite{brown93}.} a group of infinitesimal
transformations related to finite arbitrariness arising in
$S$--matrix elements upon elimination of ultraviolet (UV)
divergences. These authors used the notion of {\it normalization
group} as a Lie transformation group generated by differential
operators connected with renormalization of the coupling constant
$e$.

 In the following year, on the basis of (infinite) Dyson's renormalization
transformations formulated in the regularized form, Gell-Mann and
Low \cite{gml} derived functional equations (FEs) for the QED
propagators in the UV limit. The appendix to this article
contained the general solution (obtained by T.D. Lee) of FE for
the renormalized transverse photon propagator amplitude
$d(Q^2/\lambda^2, e^2)$ ($\lambda$ -- cutoff defined as a
normalization momentum). This solution was used for a qualitative
analysis of the quantum electromagnetic interaction behaviour at
small distances. Two possibilities, namely, infinite and finite
charge renormalizations were pointed out. However,
paper~\cite{gml} paid no attention to the group character of the
analysis and of the qualitative results obtained there. The
authors missed a chance to establish a connection between their
results and the QED perturbation theory and did not discuss the
possibility that a ghost pole solution might exist.

  The decisive step was made by Bogoliubov and the present author
\cite{bs-55a,bs-55b,sh-55} in 1955~\footnote{See also two survey
papers \cite{nc-56} published in English in 1956.}. Using the
group properties of finite Dyson transformations for the coupling
constant, fields and Green functions, they derived functional
group equations for the renormalized propagators and vertices in
QED in the general (i.e., with the electron mass taken into
account) case.

 In the modern notation, the first equation
\begin{equation}    \label{1-1}
\albar(x,y;\alpha)=\albar\left(\frac{x}{t}, \frac{y}{t};
~\albar(t,y;\alpha) \right)~; ~~
x=\frac{Q^2}{\mu^2}~,~~y=\frac{m^2}{\mu^2}~,    \end{equation} is
that for the invariant charge (now widely known also as an
effective or running coupling) $~\bar\alpha=\alpha
d(x,y;\alpha=e^2)$ and the second ---
\begin{equation}    \label{1-2}
 s(x,y;\alpha)=s(t,y;\alpha)\,s\left(\frac{x}{t}, \frac{y}{t}; ~
\albar(t,y;\alpha)\right)~ \end{equation}
--- for the electron propagator amplitude.

 These equations obey a remarkable property: the product $e^2d\equiv\albar\,$
of the electron charge squared and the photon transverse
propagator amplitude enters into both FEs. This product is
invariant with respect to finite Dyson's transformation (as it is
stated by Eq.(\ref{1-1})) which now can be written in the form
\begin{equation} \label{rgt}
R_t : ~\left\{\mu^2\to t\mu^2, \alpha\to\albar(t,y;
\alpha)\right\}~. \eeq

 We called this product \textit{invariant charge} and introduced the term
\textit{ renormalization group}. \par

 Let us emphasize that, unlike in Refs.\cite{stp,gml}, in the Bogoliubov
formulation there is no reference to UV divergences and their
subtraction or regularization. At the same time, technically,
there is no simplification due to the massless nature of the UV
asymptotics. Here, the homogeneity of the transfer momentum scale
$Q$ is explicitly violated by the mass $m$. Nevertheless, the
symmetry with respect to transformation $R_t$ (even though a bit
more involved) underlying RG is formulated as an {\it exact
property} of the solution. This is what we mean when using the
term {\sf Bogoliubov renormalization group} or
\textit{renormgroup} for short.
\smallskip

 The differential Lie equations for $\albar$ and for the electron propagator
\begin{equation}\label{1-4}
\frac{\partial\albar(x,y;\alpha)}{\partial\ln
x}=\beta\left(\frac{y}{x},\,
\albar(x,y;\alpha)\right)\,\,;\,~\,~\frac{\partial
s(x,y;\alpha)}{\partial \ln
x}=\gamma\left(\frac{y}{x},\,\albar(x,y;\alpha)\right)s(x,y;\alpha)~,
\end{equation}
with
\begin{equation} \label{1-5}
\beta(y,\alpha)=\frac{\partial\bar\alpha(\xi,y;\alpha)}{\partial\xi}~,~
~~~\gamma(y,\alpha)=\frac{\partial
s(\xi,y;\alpha)}{\partial\xi}~~~~ \mbox{at}~~\xi=1~ \end{equation}
were first derived in \cite{bs-55a} by differentiating FEs
(\ref{1-1}) and (\ref{1-2}) over $x$ at the point $t=x$. On the
other hand, by differentiating the same equations over $t$ one
obtains \cite{bereg}
 \beq\label{bereg1}
 X\,\albar(x,y;\alpha)=0~;~~X\,s(x,y;
 \alpha)=\gamma(y,\alpha)s(x,y;\alpha)\eeq
with
 \beq\label{bereg2}
 X= x\partial_x  +y\partial_y-\beta(y,\alpha)\partial_{\alpha}~
 \qquad  \left(\partial_x\equiv \partial/\partial x\right)~, \eeq
the Lie infinitesimal operator.

\subsubsection{\protect\small\it{Creation of the RG method}
\label{sss1.1.2}}

 Another important achievement of \cite{bs-55a} consisted in formulating a
simple algorithm for improving an approximate perturbative
solution by combining it with Lie group equations -- for detail,
see below Section \ref{ss1.3}.

 In our adjacent publication~\cite{bs-55b} this algorithm was effectively
used to analyse the UV and infrared (IR) behaviour in QED. In
particular, the one-loop UV asymptotics of the photon propagator
as well as the IR behavior of the electron propagator in the
transverse gauge
\begin{equation} \label{a1rg}
\albar^{(1)}_{\rm
rg}(x;\alpha)=\frac{\alpha}{1-\frac{\alpha}{3\pi}\ln x}\,\,~,~~
s(x,y;\alpha)\approx (p^2/m^2-1)^{-3\alpha/2\pi}~\eeq were
derived. At that time, these expressions, summing the {\it leading
log's terms} were already known from papers by Landau with
collaborators \cite{lakh}.  However, Landau's approach did not
provide a means for constructing subsequent approximations.

  The simple technique for calculating higher approximations was found only
within the new renormgroup method. In the same paper, starting
with the next order perturbation expression $\bar\alpha^{(2)}_{\rm
pt}(x;\alpha)$ containing the $\alpha^3\ln x$ term, we arrived at
the second renormgroup approximation (see below Section
\ref{sss1.3.2})
\begin{equation}   \label{a2rg}
\bar\alpha^{(2)}_{\rm
rg}(x;\alpha)=\frac{\alpha}{1-\frac{\alpha}{3\pi}\ln x
+\frac{3\alpha}{4\pi}\ln(1-\frac{\alpha}{3\pi}\ln x)}
\end{equation} which performs infinite summation of the
$\alpha^2(\alpha\ln)^n$ terms. This two-loop solution for
invariant coupling first obtained in~\cite{bs-55b} contains the
nontrivial log--of--log dependence which is now widely known of
the ``next-to-leading logs" approximation for the running coupling
in quantum chromodynamics (QCD) --- see, below, Eq.(\ref{2-23}).

 Comparing (\ref{a2rg}) with (\ref{a1rg}), one concludes that two-loop
correction is essential in the vicinity of the ghost pole at
$\,x_1= \exp{(3\pi/\alpha)}$. This also shows that the RG method
is a regular procedure, within which it is easy to estimate the
range of applicability of its results.       \par
  Quite soon, this approach was formulated~\cite{sh-55} for the case of QFT
with two coupling constants. To the system of FEs for two
invariant couplings there corresponds a coupled system of
nonlinear differential equations (DEs). The last was used
\cite{ilya} to study the UV behavior of the $\pi-N$ interaction at
the one-loop level.
\par
 Thus, in Refs.\cite{bs-55a,bs-55b,sh-55} and \cite{ilya} RG was directly
connected with practical computations of the UV and IR
asymptotics.  Since then, this technique, the {\sl renormalization
group method} (RGM)\footnote{Being summarized in the special
chapter of the first edition of monograph \cite{Book}.}, has
become the sole means of asymptotic analysis in local QFT.

\subsection{The Bogoliubov RG: Symmetry of a solution \label{ss1.2}}

{\sf The RG transformation.} Generally, RG can be defined as a
continuous one-parameter group of specific transformations of a
partial solution (or the solution characteristic) of a problem, a
solution that is fixed by a boundary condition. The RG
transformation involves boundary condition parameters and
corresponds to some change in the way of imposing this condition.
\par

 For illustration, imagine one-argument solution characteristic $f(x)$
that has to be specified by the boundary condition $f(x_0)=f_0$.
Formally, represent a given characteristic of a partial solution
as a function of boundary parameters as well: $f(x)=f(x,x_0,f_0)$.
This step can be treated as an {\it embedding} operation. Without
loss of generality $\,f\,$ can be written in a form of a
two-argument function $F(x/x_0, f_0)$ with the property
$~F(1,\gamma)=\gamma$.

  The RG transformation then corresponds to a changeover of the way of
parameterization, say from $\{x_0,f_0\}$ to $\{x_1,f_1\}$ for the
{\it same} solution. In other words, the $x$ argument value, at
which the boundary condition is given, can be changed for $x_1$
with $f(x_1)=f_1$. The equality $F({x/x_0}, f_0)=F({x/x_1}, f_1)$
now reflects the fact that under such a change the form of the
function $F$ itself is not modified. Noting that $~f_1= F(x_1/x_0,
f_0)~$, we get
 \[ F(\xi, f_0) = F(\xi/t, F(t,f_0))~~;~~~
 \xi = x/x_0~, ~t = x_1/x_0~. \]
The group transformation here is $\{~\xi\to\xi/t,~~f_0 \to
F(t,f_0)~\}~$.

{\sf The renormgroup transformation} for a given solution of some
physical problem in the simplest case can now be defined as \par
\smallskip

{\it a simultaneous one-parameter transformation of two
variables}, say $x$ and $g$, by
 \beq\label{rgt-m0} R_t~~:~~\{~x\to
 x' = x/t~,~~g\to g'=\gbar(t,g)~ \}~,\eeq
the first being a scaling of a coordinate $x$ (or reference point)
and the second
--- a more complicated functional transformation of the solution
characteristic. The equation
\begin{equation}\label{1_9}
\bar{g}(x,g) =\bar{g}\left(~x/t\,,~\bar{g}(t,g)\right)~ \eeq for
the transformation function $\gbar$ provides the group property
$T_{\tau t}=T_\tau T_t~$ of the transformation (\ref{rgt-m0}).
\par

 They are just the RG FEs and transformation for a massless QFT model
with one coupling constant $g$. In that case $x =Q^{2}/\mu^{2}$ is
the ratio of a 4-momentum $Q$ squared to a ``normalization"
momentum $\mu $ squared and $g$, the coupling constant.  \par
  The RG transformation (\ref{rgt-m0}) of a QFT amplitude $\,s\,$ is of the
form (compare with Eq.(\ref{1-2}))
 \beq \label{rgt-s}
 R_t\,\cdot s(x,g)\equiv e^{- \ln t\, X} s(x,g)
 = s(x/t,~\bar{g}(t,g))=z^{-1}_s\, s(x,g);\ \ \
 z_s=s(t,g)\,.
\eeq

Several generalizations are in order.

\smallskip

 {\sf a. ``Massive" case.} For example, in QFT, if we do not neglect mass
$m$ of a particle, we have to insert an additional dimensionless
argument into the invariant coupling $\bar{g}$ which now has to be
considered as a function of three variables:
$x=Q^{2}/\mu^{2},~y=m^{2}/\mu^{2}$, and $g$. The presence of a new
``mass" argument $y$ modifies the group transformation
(\ref{rgt-m0}) and the FE (\ref{1_9}) \beq\label{1-12}
R_t\/:~\left\{x'=\frac{x}{t}\,,~y'=\frac{y}{t}\,,~g'=\bar{g}(t,y;g)~\right\}\,
;~~
\bar{g}(x,y;g)=\bar{g}\left(~\frac{x}{t}~,~\frac{y}{t}\,;~\bar{g}(t,y;g)
\right)~. \eeq
 Here, it is important that the new parameter $y$ (which, physically,
should be close to the $x$ variable, as it scales similarly)
enters also into the transformation law of $g$ . \par

If the considered QFT model, like QCD, contains several masses,
there will be several mass arguments $y\to\{y\}\equiv y_1, y_2,
\ldots, y_n~.$

\smallskip

 {\sf b. Multi-coupling case.} A more involved generalization corresponds
to transition to the case with several coupling constants:
$g\to\{g\}=g_1, \ldots, g_k~$. Here, there arise a ``family" of
effective couplings \beq\label{1-15}
\bar{g}\to\{\bar{g}\}~,~~\bar{g}_i=\bar{g}_i(x,y;\{g\})~;~~~~~i=1,2,\ldots,
k~, \eeq \noindent satisfying the system of coupled functional
equations \beq\label{1-16}
\bar{g}_{i}(x,y;\{g\})=\bar{g}_i\left(x/t\,,~y/t~;~ \{
~\bar{g}(t,y; \{g\} )~\}~\right)~. \eeq
 The RG transformation now is
\beq\label{1-17} R_t~:~\left\{~x \to x/t\,,~~y \to y/t~,~~\{g\}
\to \{g(t)\}~\right\}~,~~~ g_i(t)=\gbar_i(t,y;\{g\})~.  \eeq

\subsection{Renorm-group method \label{ss1.3}}

\subsubsection{\protect\small\it{The algorithm} \label{sss1.3.1}}

  The idea of the approximate solution marriage~\cite{bs-55a,bs-55b} to group
symmetry can be realized with the help of group DEs. If we define
$\,\beta$ and $\gamma\,$ (the so-called ``generators" on physical
slang) from some approximate solutions and then solve evolutional
DEs, we obtain the {\it RG improved} solutions that obey the group
symmetry and correspond to the approximate solutions used as an
input.
\smallskip

   Now we can formulate an algorithm of improving an approximate solution.
The procedure is given by the following prescription which we
illustrate by a massless one--coupling case (\ref{1-4}) and
(\ref{1-5}): \par
  Assume some approximate solution $\,\gbar_{\rm appr}(x, g)\,, s_{\rm appr}
(x, g)\,$ is known.

{\bf 1.} On the basis of Eq.(\ref{1-5}) define the beta-- and
gamma--functions \beq\label{2-18} \beta (g)\define
{\partial\over{\partial\xi}}\gbar_{\rm appr}(\xi, g) \bigg|_{\xi =
1} ~;~~~\gamma(g)\define {\partial\over{\partial\xi}} s_{\rm
appr}(\xi, g) \bigg|_{\xi = 1} ~.  \eeq

{\bf 2.} Integrate the first of Eqs.(\ref{1-4}), i.e., construct
the function \beq\label{2-19}
f(g) \define\int\nolimits^g{{d \tau}\over{\beta (\tau)}}~,\eeq %

{\bf 3.} Resolve the obtained equation, i.e., \beq\label{2-20}
 \gbar_{\rm rg} (x, g)= f^{-1} \{ f(g) + \ln x \} ~~. \eeq

{\bf 4.} Integrate the second of Eqs.(\ref{1-4}) using this
expression $\,\gbar_{\rm rg}\,$ in its r.h.s. to obtain $\,s_{\rm
rg}(x,g)\,$ in the explicit form.

{\bf 5.} Then, the expressions $\,\gbar_{\rm rg}\,$ and $\,s_{\rm
rg}\,$ precisely satisfy the RG symmetry, i.e., they are exact
solutions of Eqs.(\ref{1_9}) and (\ref{rgt-s}) corresponding to
$\,\gbar_{\rm appr}\,$ and $\,s_{\rm appr}\,$ used as an input.

\subsubsection{\protect\small\it{Simple illustration} \label{sss1.3.2}}

For illustration, take the simplest perturbative expression
$\gbar_{\rm pt}^{(1)}=g-g^2\beta_1\ln x $ for $\,\gbar_{\rm
appr}\,$ and
 $\,s_{\rm pt}^{(1)}=1-g\gamma_1\ln x $. Here, $\beta(g)=-\beta_1 g^2,
~\gamma(g)=-\gamma_1 g\,$ and integration of (\ref{1-4}) gives
explicit expressions
\begin{equation}\label{2-21}
\gbar_{\rm rg}^{(1)}(x,g)={g\over 1+g\beta_1\ln x}~,~~~s_{\rm
rg}^{(1)}(x,g)=
\left(\gbar(x,g)/g\right)^{\nu_1}~;~~~\nu_1=\gamma_1/\beta_1\,,
\eeq which, on the one hand, exactly satisfy the RG symmetry and,
on the other, being expanded in powers of $g$, correlate with
$\gbar_{\rm pt}$ and $s_{\rm pt}$.

Now, on the basis of geometric progression (\ref{2-21}), let us
present the 2-loop perturbative approximation for $\,\gbar\,$ in
the form $\gbar_{\rm pt}^{(2)}=g-g^2\beta_1\ln x+g^3
(\beta_1^2\ln^2 x-\beta_2\ln x).$ By using this expression as an
input in Eq.(\ref{2-18}), we have $ \beta^{(2)}(g)= -\beta_1
g^2-\beta_2 g^3 $ and then (step {\bf 2})
 $$ \beta_1f^{(2)}(z)= -\int^z{d\tau \over\tau^2+b\tau^3}=
 {1\over z}+b \ln {z \over1+ b z}~;~~~b={ \beta_2 \over \beta_1}~.
 $$

 To make the last step, we have to start with the equation
$$ f^{(2)}[\gbar^{(2)}_{\rm rg}(x,g)]=f^{(2)}(g)+ \beta_1\ln x $$
which is a transcendental one and has no simple explicit
solution\footnote{It can be expressed via special, Lambert,
$W$-function: $W(z)\exp^{W(z)}=z$; see, e.g.,
Ref.\cite{s-tmf99}.}. Due to this, one usually resolves this
relation approximately. Take into account that the second,
logarithmic, contribution to $f^{(2)}(z)$ is a small correction to
the first one at $bz\ll 1$. Under this reservation, we can
substitute the one-loop RG expression (\ref{2-21}) instead of
$\gbar^{(2)}_{\rm rg}$ into this correction and obtain the
explicit ``iterative" solution \beq\label{2-23} \gbar_{\rm
rg}^{(2)}= {g\over{1 + g\beta_1 l +
g(\beta_2/\beta_1)\ln\left[1+g\beta_1l\right]}} ~;~~~~ l = \ln x.
\eeq

An analogous procedure for $ s_{\rm pt}^{(2)}=1-g\gamma_1\ln x+
g^2\left(\gamma_1(\gamma_1+1)\beta^2_1\ln^2 x-\gamma_2\ln x\right)
$ yields
 \beq\label{s-rg2}
 s_{\rm rg}^{(2)}= \frac{S\left(\gbar_{\rm rg}^{(2)}(x, g)\right)}{S(g)}~~~
 \mbox{with}~~~S(g)=g^{\nu_1}e^{\nu_2g}~~\mbox{and}~~
 ~~~\nu_2=\frac{\beta_1\gamma_2-\gamma_1\beta_2}{\beta_1^2}~. \eeq

These results are interesting from several aspects. \par
\begin{itemize}
\item
First, being expanded in $g$ and $gl$ powers, they produce an infinite
series containing ``leading" and ``next-to-leading" UV logarithmic
contributions.
\item
Second, they contain a nontrivial analytic dependence $\ln (1+g\beta_1l)
\sim \ln(\ln Q^2)$ which is absent in the perturbation input.
\item
Third, being compared with the one-loop solution, Eq.(\ref{2-21}),
they demonstrate an algorithm of subsequent improving of accuracy,
i.e., of RGM regularity.
\end{itemize}

\subsubsection{\protect\small\it{RGM usage in QFT}} \label{sss1.3.3}

 As we have seen, QFT perturbation expression of finite order does not obey
the RG symmetry. On the other hand, it was shown that the one-loop
and two-loop approximations, used as an input for the construction
of ``generators" $\beta(g)$ and $\gamma(g)$, yield expressions
(\ref{2-21}), (\ref{2-23}) and (\ref{s-rg2}) that obeys the group
symmetry and exactly satisfy FEs (\ref{1_9}) and (\ref{rgt-s}).
\par

   More generally, one can state the following logical structure of the
RGM procedure. \par
--- Solving group equation(s) for invariant coupling(s) $\gbar_{\rm rg}(x,g)$
using some approximate solution $\gbar_{\rm pt}$ as an input. \par
--- Obtaining RG solutions for some other QFT objects (like vertices and
propagator amplitudes) on the basis of the expression(s) for
$\gbar_{\rm rg}$ just derived. Typically, they satisfy the
equation
\begin{equation}
X\, M(x,y,g)=\gamma(y,g)\, M(x,y,g)\,.   \end{equation}

General structure of the corresponding solutions has the form
\beq\label{genRGsol} M(x,y; g)=z^{-1}_M(y,g)\,{\cal
M}\left(x/y\,,~ \gbar(x,y; g)\right).\eeq
 Note that the function ${\cal M}$ in the r.h.s. depends only on the RG
invariants, that is on the first integrals of the RG operator $X$
introduced in Eqs.(\ref{bereg1}) and (\ref{bereg2}). It satisfies
homogeneous partial differential equations (PDEs) $X\, {\cal
M}=0$. For the RG invariant objects, like observables, $z_M=1$,
$\gamma = 0$.
\smallskip

Now we can resume the RGM properties. The RGM is a regular
procedure of {\it combining} dynamical information (taken from an
approximate solution) with the RG symmetry. The essence of RGM is
the following: \smallskip

\noindent 1) The mathematical tool used in RGM is Lie differential
equations.          \smallskip

\noindent 2) The key element of RGM is possibility of an
(approximate) determination of ``generators", like $\beta(g),
\gamma(g)$, from dynamics.
 \smallskip

\noindent 3) The RGM works effectively in the case when the
solution has a singular behaviour. It restores the structure of
singularity compatible with the RG symmetry.

\section{Evolution of Renormalization Group \label{s2}}

  In the 70s and 80s RG ideas have been applied to critical phenomena:
spontaneous magnetization, polymerization, percolation,
non-coherent radiation transfer, dynamic chaos, and so on. Less
sophisticated motivation by Wilson in spin lattice phenomena (than
in QFT) made this ``explosion" of RG applications possible.

\subsection{Renormalization Group Evolving \label{ss2.1}}

\subsubsection{\protect\small\it{Kadanoff--Wilson RG in critical phenomena}}

{\sf a. Spin lattice.} The so--called {\it renormalization group
in critical phenomena} is based on the Kadanoff--Wilson procedure
\cite{leo,ken} referred to as ``decimation" or ``blocking".
Initially, it emerged from the problem of spin lattice. \par
 Imagine a regular (two-- or three--dimensional) lattice consisting of
$N^d, ~d=2,3$ cites with an `elementary step' $a$ between them.
Suppose that at every site a spin vector $\bm{\sigma}$ is located.
The Hamiltonian, describing the spin interaction between nearest
neighbours
 \[ H = k \sum_i \bm{\sigma}_i\cdot \bm{\sigma}_{i \pm 1} \]
 contains $k$, the coupling constant. A statistical sum is obtained from
the partition function, $\,S=<\exp(-H/\theta)>_{\rm aver}.$

 To realize blocking, one has to perform the ``spin averaging" over block
consisting of $n^d$ elementary sites. This step diminishes the
number of degree of freedom from $N^d$ to $(N/n)^d$. It also
destroys the small-range properties of a system, in the averaging
course some information being lost. However, the long-range
physics (like correlation length essential for phase transition)
is not affected by it, and we gain simplification of the problem.

  As a result of this blocking procedure, new effective spins $\bm{\Sigma}$
arise in new sites forming a new effective lattice with a step
$na$. We arrive also at the new effective Hamiltonian
 \[  H_{\rm eff}=K_n \sum_I \bm{\Sigma}_I
    \cdot\bm{\Sigma}_{I\pm1}+\Delta H~,  \]
with the effective coupling $K_n$ between new spins
$\bm{\Sigma}_I$ of new neighbouring sites; $K_n$  has to be
defined by the averaging process as a function of $\,k\,$ and
$\,n\,$. Here, $\Delta H$ contains quartic and higher spin forms
which are irrelevant for the IR (long-distance) properties. Due to
this, one can drop $\,\Delta H\,$ and conclude that the spin
averaging leads to an approximate transformation,
 \[ k \sum_i \bm{\sigma} \cdot \bm{\sigma} \to
 K_n \sum_I \bm{\Sigma} \cdot \bm{\Sigma} ~~, \]
or, taking into account the ``elementary step" change, to $
\left\{a \to n\,a, ~k \to K_n \right\}$. The latter is the
Kadanoff--Wilson transformation. It is convenient to write down
the new coupling $K_n$ in the form $K_n= K(1/n,K)$. Then, the KW
transformation reads \beq\label{kw} KW_n: \left\{a \to n a, \quad
k\to K_n=K\left({1/n}, k\right)\right\}~.\eeq These
transformations obey composition law $KW_n\cdot KW_m=KW_{nm}$ if
the relation \beq K(x, k) = K({x/t}, K(t, k))\,, \quad x={1/nm},
\quad t= {1/n}~.\eeq holds. This is very close to RG symmetry.
\smallskip

 We observe the following points:
\begin{itemize}
\item The RG symmetry is approximate (due to neglecting $\Delta H$).
\item The transformations $KW_n\,$ are discrete.
\item There exist no reverse transformation to $KW_n\,$.
\item Transformations $\,KW_n\,$ relate different auxiliary models.
\end{itemize}
 Hence, the `Kadanoff--Wilson renormalization group` (KW--RG) is an {\it
approximate and discrete semi-group}. For a long--distance (IR
limit) physics, however, $\Delta(1/n)$ is small and it is possible
to use differential Lie equations\footnote{In application of
these transformations to critical phenomena, the notion of a {\it
fixed point} is important. Generally, a fixed point is associated
with power-type asymptotic behavior. Note here that, contrary to
the QFT case considered in Section \ref{sss1.3.2}, in phase
transitions we deal with the IR stable point.}.

{\sf b. Polymer theory.} In polymer physics, one considers
statistical properties of polymer macromolecules which can be
imagined as a very long chain of identical elements (with the
number of elements $N$ as big as $10^5$). Molecules are swimming
in a solvent and form {\it globulars}.  This big molecular chain
forms a specific pattern resembling the pattern of a random walk.
The central problem of the polymer theory is very close to that of
a random walk and can be formulated as follows. \par

 For a long chain of $N$ ``steps" (the size of step = $a$), one has to find
the ``chain size" $R_N$, the distance between the ``start" and the
``finish" points (the size a of globular), with the distribution
function $f(\phi)$ of angles between the neighboring elements
being given. \par

  For large $N$ values, the molecular size $R_N$ follows the power {\it
Fleury law} $R_N \sim N^\nu $ with $\nu$, the Fleury index. When
$N$ is given, $R_N$ is a functional of $f(\phi)$ which depends on
external conditions (e.g., temperature $T$, properties of solvent,
\etc). If $T\/$ grows, $R_N\/$ increases and at some moment
globulars touch one another. This is the polymerization process
which is very similar to a phase transition phenomenon. \par
   The Kadanoff--Wilson blocking ideology has been introduced in physics of
polymers by De Gennes \cite{degen}. The key idea is a grouping of
$n$ neigbouring elements of a chain into a new ``elementary
block". It leads to the transformation $\left\{1\to n~;~~a \to A_n
\right\}~ $ which is analogous to one for the spin lattice
decimation. This transformation must be specified by a direct
calculation which gives an explicit form of $A_n=\bar{a}(n,a)$.
Here, we have a discrete semi-group. Then, by using the KW--RG
technique, one finds the fixed point, obtains the Fleury power law
and can calculate its index $\nu$.

 The essential feature of a polymer chain is the impossibility of a
self-intersection. This is known as an {\it excluded volume}
effect in the random walk problem.  Generally, the excluded volume
effect yields some complications. However, inside the alternative,
the QFT RG approach to polymers \cite{alhim}, it can be treated
rather simply by introducing an one more argument which is similar
to finite length $L$ in the transfer problem or particle mass $m$
in QFT.  \smallskip

 Besides polymers, the KW--RG technique has been used in some fields of
physics, like percolation, non-coherent radiation transfer
\cite{bell}, dynamical chaos \cite{chaos} and some others.
\smallskip

\subsubsection{\protect\small\it{Bogoliubov symmetry outside QFT}}
\label{sss2.1.2}

 Meanwhile, the original QFT--RG approach proliferated into some other parts
of theoretical physics. In the late 50s, it was used~\cite{jetp59}
for summation of Coulomb singularities in Bogoliubov's theory of
superconductivity based on the Fr\"ohlich electron--phonon
interaction. Twenty years later it was used in the theory of
turbulence. \par

{\sf a. Turbulence.} To formulate the turbulence problem in terms
of RG, one has to perform the following steps ~\cite{dom,vas}:
\begin{enumerate}
\item Introduce the generating functional for correlation functions.
\item Write down the path integral representation for this functional.
\item By changing the functional integration variable, find the equivalence
     of the statistical system to some quantum field theory model.
\item Construct the system of Schwinger--Dyson equations for this equivalent
     QFT model.
\item
Perform the finite renormalization procedure and derive the RG equations.
\end{enumerate}
\smallskip

 Here, the reparameterization degree of freedom, physically corresponds to a
change of long wave-length cutoff which is built-in into the
definition of a few effective parameters. \smallskip

  {\sf b. Weak shock wave.} Another example can be taken from hydrodynamics.
Consider a weak shock wave in the one-dimensional case of a large
distance $l$ from the starting (implosion) point. The dependence
of velocity $v$ of a matter as a function of $l$ at a given moment
of time $t$ has a simple triangular shape and can be described by
the expression
 \[ v(l)=\frac{l}{L} V~\mbox{at}~~~~l \leq  L~;\qquad
          = 0 \ \mbox{for} \quad l>~L\,,
 \]
where $L=L(t)$ is the front position and $V=v(L)$ -- the front
velocity. They are functions of time. In the absence of viscosity,
the ``conservation law" $~LV = {\rm Const.}~$ holds. Due to this,
they can be treated as functions of the front wave position
$L\equiv x, V=V(x)$ as well. If the physical situation is
homogeneous, then the front velocity $V(x)$ should be considered
as a function of only two additional relevant arguments -- its own
value $V_0 =V(x_0)$ at some precedent point $(x_0 < x)~$  and of
the $x_0$ coordinate. In can be written down in the form :
$~V(x)=G(x/x_0, V_0)~$. If we pick up three points $x_0$, $x_1$
and $x_2$ (for details, see Refs.\cite{fss84,O-88}), then the
initial condition may be given either at $x_0$ or $x_1$. Thus, we
obtain the FE equation equivalent to (\ref{1_9}) $$ V_2 =
G(x_2/x_0, V_0) = G(x_2/x_1, V_1)=G(x_2/x_1, G(x_1/x_0, V_0))~.$$

{\sf c. One-dimensional transfer.}
 A similar argument has been done by Mnatzakanian~\cite{mamik} in the
transfer problem at one dimension. Imagine a half--space filled
with a homogeneous medium on the surface of which some flow (of
radiation or particles) with intensity~$g_0$ falls from the vacuum
half--space. \par
    Follow the flow as it moves inwards the medium at the distance~$l$
from the boundary. Due to homogeneity along the $l$ coordinate,
the intensity of the penetrated flow $g(l)$ depends on two
essential arguments, $g(l)=G(l,g_0)$. The values of the flow at
three different points $g_0$  (on the boundary), $g_1$ and $g_2 >
g_1$ can be connected with each other by the transitivity
relations, $g_1 =G(\lambda, g_0)\,, \quad g_2=G(\lambda+l,
g_0)=G(l, g_1)\,,$ which lead to the FE
\begin{equation}
\label{add-feq} G(l, g)=G(l-\lambda, G(\lambda, g))\,.
\end{equation} \par Performing a logarithmic change of variables $
 l =\ln x,~\lambda =\ln t,~ G(l,g) = \bar{g}(x,g)~$,
we see that (\ref{add-feq}) is equivalent to (\ref{1_9}).
\smallskip

 Consider now intensity of a reverse flow, that is total amount of particles
at the point $l$ moving in the backward direction. It is
completely defined by $g_0$ and can be written down as $R(l,
g_0)$. This function can be represented in the form $\,R(l, g) =
R_0(g) N(l, g)\,$ with $\,R_0 \equiv R(0, g)\,$ and function $N$
``normalized" on the boundary $N(0, g)=1$. Playing the same game
with transitivity, we arrive at FE
\begin{equation} \label{n2}
   N(l, g)=Z(l, g) N(l-\lambda, G(l, G(\lambda, g))~; \quad
   Z= R_0(g_1)/R_0(g)
\end{equation}
related to Eq.(\ref{rgt-s}) by logarithmic change of variables.
One can refer to (\ref{add-feq}) and (\ref{n2}) as to the {\it
additive version} of RG FEs and to previous equations of Section
1, like (\ref{1_9}), (\ref{rgt-s}) and (\ref{1-12}) as to the {\it
multiplicative} one.
\smallskip

  The transfer problem admits a modification connected with discrete
inhomogeneity: imagine the case of two different kinds of
homogeneous materials separated by the inner boundary surface at
$\,l=L$. The point of breaking $~l = L~$ may correspond to the
boundary with empty space, and resulting equation is equivalent to
Eq.(\ref{1-12}). \par
  One more generalization is related to ``multiplication" of argument $g$ as
expressed by Eq.(\ref{1-15}). Physically, this relates to the case
of radiation on different frequencies $\omega_i\,,~
i=1,2,\dots\,k\,$ (or particles of different energies or of
different types).\par
  Take the case of $k=2$ and suppose that the material of the medium has such
properties that the transfer processes of the two flows are not
independent. In this case, the characteristic functions of these
flows $G$ and $H$ are dependent on both the boundary values $g_0$
and $h_0$ and can be taken as functions $g(l) = G(l, g_0, h_0)~, \
\ h(l) = H(l, g_0, h_0)~$. \par

After a group operation $~l \to l - \lambda~$, we arrive at a
coupled set of functional equations
 $$ G(l + \lambda,g,h)= G(l,g_\lambda,h_\lambda)\,,~
    H(l+\lambda,g,h)= H(l, g_\lambda,h_\lambda)\,;~
    g_\lambda\equiv G(\lambda,g,h)\,,~
    h_\lambda \equiv H(\lambda, g,h)
 $$
which is just an additive version of system (\ref{1-16}) at $k=2$.
\par
\smallskip

Now we can make the important conclusion that a common property
yielding functional group equations is just the transitivity
property of some physical quantity with respect to the way of
giving its boundary or initial value. \par

Hence, the RG  symmetry is not a symmetry of equations but a
symmetry of solution, that is of equations and boundary conditions
considered as a whole.

\subsection{Difference between Bogoliubov RG and KW--RG \label{2.2}}

 As we have mentioned above, the RG ideas expanded in diverse fields of
physics in two different ways:
\begin{itemize}
\item via direct analogy with the Kadanoff--Wilson construction (averaging
over some set of degrees of freedom) in polymers, non-coherent
transfer and percolation, i.e., constructing a set of models for a
given physical problem.
\item via finding an exact RG symmetry by proof of the equivalence with a
QFT model (e.g., in turbulence \cite{dom,vas}), plasma turbulence
\cite{pell}) or by some other reasoning (like in a transfer
problem).
\end{itemize} \bigskip

 To the question {\sf Are there different renormalization groups?}  the
 answer is positive:
\begin{enumerate}
\item In QFT and some simple macroscopic examples, ~RG~ symmetry is an exact
symmetry of the solution formulated in its natural variables.
\item In turbulence, continuous spin-field models and some others, it
is a symmetry of an equivalent QFT model.
\item In polymers, percolation, \etc, (with KW blocking), the RG
transformation is a {\it transformation between different
auxiliary models} (specially constructed for this purpose) of a
given system.
\end{enumerate}
\medskip

 As we have shown, there is no essential difference in the mathematical
formulation. There exists, however, a profound difference in
physics:
\par --- In the cases 1 and 2 (as well as in some macroscopic examples), the RG
is an exact symmetry of a solution. \par
 --- In the Kadanoff--Wilson type problem (spin lattice, polymers,
\etc), one has to construct a set ${\cal M}$ of models $M_i$. The
KW--RG transformation
 \beq KW_{n}\, M_i = M_{n i} ~,~~\mbox{with integer} ~n \eeq
 {\it is acting inside a set of models}.

\subsection{Functional self-similarity} \label{ss2.4}

 The RG transformations have close connection with the
concept of self-similarity. The self-similarity transformations
for problems formulated by nonlinear PDEs are well known since the
last century, mainly in dynamics of liquids and gases. They are
one parameter $\lambda$ transformations defined as a simultaneous
power scaling of independent variables $~z=\{x,t,\ldots\}~$,
solutions $~f_k(z)~$ and other functions $~V_i(z)~$ (like external
force) $$ S_\lambda : ~\left\{ x^\prime = x \lambda\,,~t^\prime =
t\lambda^a\,,~ f_k^\prime =\lambda^{\varphi_k} f_k~,~ V_i^\prime
=\lambda^{\nu_i} V_i~\right\} ~~ $$ entering into the equations.
\par

To emphasize their power structure, we use a term {\it power}
self-similarity = PS. According to Zel'dovich and Barenblatt,
\cite{zeld} the PS can be classified as: \par

a/ {\it PS of the 1st kind} \ \ with all indices $a, ... \varphi,
\nu , ...$ being integers or rational (Rational PS) that are
usually found from the theory of dimensions; \par b/ {\it PS of
the 2nd kind} \ \ with irrational indices (Fractal PS) which
should be defined from dynamics.

 To relate RG with PS, turn to the renormgroup FE $\gbar(xt, g)=\gbar (x,
\gbar(t, g))$.  Its general solution is known; it depends on an
arbitrary function of one argument -- see Eq.(\ref{2-20}).
However, at the moment, we are interested in a special solution
linear in the second argument:  $\gbar (x, g)= g X(x).$ The
function $X(x)$ should satisfy the equation $~X(xt)=X(x)X(t)~$
with the solution $~X(x)=x^{\nu}~$. Hence, $\gbar (x, t) = g x^\nu
$. This means that in our special case, linear in $g$, the RG
transformation (\ref{rgt-m0}) is reduced to the PS transformation,
 \beq \label{6-1}
 R_t \quad \Rightarrow \quad S_t: \{ x^{\prime} = xt^{-1},
          ~g^{\prime} = gt^\nu \} ~~.\eeq

 Generally, in RG, instead of a power law, we have an arbitrary functional
dependence. Thus, one can consider transformations (\ref{rgt-m0}),
(\ref{1-12}) and (\ref{1-17}) as functional generalizations of
usual (i.e., power) self-similarity transformations. Hence, it is
natural to refer to them as to the transformations of {\it
functional scaling} or functional (self)similarity (FS) rather
than to RG-transformations. In short, $$ {\rm RG} \equiv {\rm FS}
~~, $$ with FS standing for {\sf Functional
Similarity}\footnote{This notion was first mentioned in
\cite{BBSh57} and formally introduced~\cite{dan82} in the
beginning of 80s.}.

  Now we can answer the question on the physical meaning of the symmetry
underlying FS and the Bogoliubov renormgroup. As we have
mentioned, it is not a symmetry of a physical system or of
equation(s) of the problem at hand, but a {\it symmetry of a
solution} considered as a function of the relevant physical
variables and suitable boundary parameters. A symmetry like that
can be related, in particular, to the invariance of a physical
quantity described by this solution with respect to the way in
which the boundary conditions are imposed. The changing of this
way constitutes a group operation in the sense that the group
composition law is related to the transitivity property of such
changes.

 Homogeneity is an important feature of a physical system under consideration.
However, homogeneity can be violated in a discrete manner. Imagine
that such a discrete violation is connected with a certain value
of $\,x\,$, say, $\,x=y$. Here, RG transformation with the
canonical parameter $\,t\,$ has the form (\ref{1-12}).

 The symmetry connected with FS is a very simple and frequently encountered
property of physical solutions. It can easily be ``discovered" in
numerous problems of theoretical physics like classical mechanics,
transfer theory, classical hydrodynamics, and so on
\cite{dan82,mamik,fss84,O-88} -- see, above, Section 2.1.2.

\section{Symmetry of solution in Mathematical Physics}
\subsection{Constructing RG-symmetries and their use}

 From the discussion in Sections 1.1 and 1.2 it follows that FS
transformation in QFT is the scaling transformation of an
independent variable $x$ (and, possibly, the parameter $y$)
accompanied by a functional transformation of the solution
characteristic $g$. It is introduced by means of either finite
transformations (\ref{rgt-m0}), (\ref{1-12}) and (\ref{1-17}) or
the infinitesimal operator (\ref{bereg2}).  Hence, the symmetry of
a solution, i.e., FS symmetry, is commonly understood in QFT as
the \textit{Lie point symmetry of a one-parameter transformation
group} defined by the operator of the (\ref{bereg2})--type. \par

Now, we are interested in getting answers to the following
questions:
\begin{itemize}
\item
is it possible to extend the notion of RG symmetry (RGS) and
generalize the form of RGS implementation that may differ from
that given by (\ref{bereg2}) ?   \hspace{2mm} --- \hspace{2mm} and
if "yes",
\item
is it possible to create a regular algorithm of finding these symmetries ?
\end{itemize}

  The answer is positive to both the questions, and below we demonstrate the
regular algorithm of constructing RGS in mathematical physics that
up to now has been devised only for boundary value problem (BVP)
for the (system of) differential equation(s) which we shall refer to as
\textit{basic equations} (BEs).  The point is that these models can be
analyzed by methods of Lie group analysis which employ
infinitesimal group transformations instead of the finite one.  \par

  The general idea of the algorithm is to find a specific {\it renormgroup
manifold} $\cal{RM}$ that contains the desired solution of BVP.
Then construction of a RGS, that leaves this solution unaltered,
is performed by standard methods of a group analysis of DEs. \par

The regular algorithm of constructing RGS (and their application)
can be formulated in a form of a scheme\footnote{In the present form
this scheme was described in \cite{KPSh-JMP98}. One can find there
historical comments and references on the pioneering publications.}
which comprises a few steps. It is illustrated in Figure~1.
 \begin{figure}[hbtp] \label{fig1}
 \centerline{\includegraphics[width=0.8\textwidth]{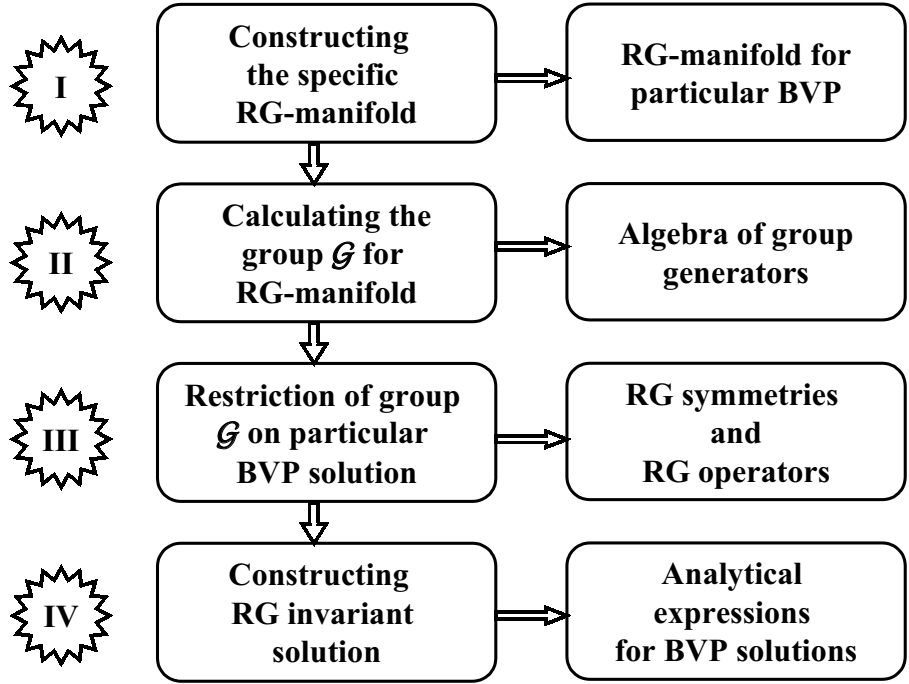}}
 \caption{RGS constructing and application to BVP in Mathematical Physics.}
 \end{figure}

\textbf{I.} \textbf{First} of all, a specific \textit{renormgroup
manifold} $\cal{RM}$ for the given BVP should be constructed which
is identified below with a system of the $k$th-order DEs
 \begin{equation}  \label{rgm}
 F_{\sigma}(z,u,u_{(1)},\ldots ,u_{(k)})=0\,,\quad \sigma=1,\ldots{},s\,.
 \end{equation}
In (\ref{rgm}) and what follows we use terminology of group
analysis and the notation of differential algebra. In contrast
with the mathematical analysis, where we usually deal with
functions $u^{\alpha},\ \alpha=1, \ldots{},m$ of independent
variables $x^i,\ i=1,~\ldots , n$ and derivatives
$u^{\alpha}_i(x)\equiv\partial u^{\alpha}/\partial
x^i,~u^{\alpha}_{ij}(x) \equiv\partial^2 u^{\alpha}/\partial
x^i\partial x^j~, \, \ldots$ that are also considered as functions of
$x$, in differential algebra we treat $u^{\alpha}, \,
u^{\alpha}_i,\, u^{\alpha}_{ij}\,,\ldots$ as variables as well.
Therefore, in differential algebra we deal with an infinite number
of variables
 \begin{equation}  \label{var}
 x=\{x^{i}\}, \quad u=\{u^{\alpha}\}\,,\quad u_{(1)}=\{u^{\alpha}_{i}\},\quad
 u_{(2)}=\{u^{\alpha}_{i_1 i_2}\} ,\, \ldots \ ; \quad
 (i,i_1,\ldots =1, \ldots , n) \,,
 \end{equation}
where $x^i$ are called \textit{independent variables},
$u^{\alpha}$ \textit{dependent variables} and $u_{(1)}, u_{(2)},
\ldots $ \textit{derivatives}. A locally analytic function $f(x,u,
u_{(1)}, \ldots ,u_{(k)} )$ of variables (\ref{var}), with the
highest $k$th-order derivative involved, is called a
\textit{differential function} of order $k$. The set of all
differential functions of a given order form a space of
differential functions $\cal A$, the \textit{universal space} of
modern group analysis \cite{Oves}--\cite{Ibr}.

 A particular way of realization of the \textbf{first step} can hardly be
described uniquely, as it depends on both a form of basic
equations and a boundary condition; generally, $\cal{RM}$ does not
coincide with BEs. We indicate here a few feasible routines for
this step.
\begin{itemize}
\item
One can use an extension of the space of variables involved in
group transformations. These variables, e.g., may be
\textit{parameters}, $p=\{ p^j \}\,, j=1, \ldots , l \/$ entering
into a solution via the equations and/or boundary conditions.
Adding parameters $p$ to the list of independent variables
$z=\{x,p\}$ we treat BEs in this extended space as $\cal {RM}$
(\ref{rgm}). Similarly, one can extend the space of differential
variables by treating derivatives with respect to $p$ as
\textit{additional} differential variables.
\item
 Another possibility employs reformulating of boundary conditions in terms
of \textit{embedding equations} or \textit{differential
constraints} which are then combined with BEs. The key idea here
is to treat the solution of BVP as an analytic function of
independent variables and boundary parameters $b=\{x_0^i,
u^{\alpha}_{0} \}$ as well. Differentiation with respect to these
parameters gives additional DEs (\textit{embedding equations})
that, together with BEs, form $\cal{RM}$. In some cases, while
calculating \textit{Lie point} RGS, the role of embedding
equations can be played by differential constraints (for details
see \cite{KPSh-JMP98}) that come from an invariance condition for
BEs with respect to the \textit{Lie-B\"acklund} \footnote{We use
here the terminology adopted in Russian literature
\cite{Oves,Ibr}. This symmetry is also known as generalized or
higher-order symmetry \cite{Olver,Sprav}.} symmetry group.
\item
 In the case when BEs contain a small parameter $\alpha$, the desired
$\cal{RM}$ can be obtained by simplification of these equations
and use of ``perturbation methods of group analysis" (see Vol.3,
Chapter 2, p.31 in \cite{Sprav}). The main idea here is to
consider a simplified ($\alpha=0$) model, which admits a wider
symmetry group (see examples in the Section \ref{ss4.2} below) in
comparison with the case $\alpha\neq0$. When we take the
contributions from small~$\alpha$ into account, this symmetry is
{\sl inherited} by BEs, which results in the additional terms,
corrections in powers of~$\alpha$, in the RGS generator.
 \end{itemize}

\textbf{II.} The \textbf{next} step consists in calculating the
most general symmetry group $\cal{G}$ that leaves the manifold
$\cal{RM}$ unaltered. The term ``symmetry group", as used in the
classical group analysis, means the property of the system
(\ref{rgm}) to admit a local Lie group of point transformations in
the space $\cal{A}$. \par
  The Lie algorithm of finding such symmetries consists in
constructing tangent vector fields defined by the operator
\begin{equation} \label{oper}
 X=\xi^{i}\partial_{x^i} + \eta^{\alpha} \partial_{u^{\alpha}}\,,
 \quad \xi^{i}\,, \eta^{\alpha} \in \cal{A}\,,
\end{equation}
with the coordinates, $\xi^{i}\,, \ \eta^{\alpha}$ that are
functions of group variables and have to be determined by a system
of equations
 \begin{equation}  \label{deteq}
  {X\, F_{\sigma}}_{\Big\vert {(\ref{rgm})}} = 0 \,, \quad
 \sigma=1,\ldots{},s\,,
 \end{equation}
that follow from the invariance of $\cal{RM}$. Here $X$ is
extended\footnote{The extending of generators to the derivatives
employs the \textit{prolongation formulas} and is a regular
procedure in group analysis (see, e.g. \cite{Sprav}).} to all
derivatives involved in $F_{\sigma}$ and the symbol
$\mid_{(\ref{rgm})}$ means calculated on the frame (\ref{rgm}). A
system of linear homogeneous PDEs (\ref{deteq}) for coordinates
$\xi^{i}\,, \ \eta^{\alpha}$, known as \textit{determining
equations}, is an overdetermined system as a rule. The solution of
Eqs.(\ref{deteq}) define a set of infinitesimal operators
(\ref{oper}) (also known as group generators), which correspond to
the admitted vector field and form a Lie algebra. In the case that
the general element of this algebra
 \begin{equation}
 \label{gen-oper}
 X = \sum\limits_{j} A^j X_j \,,
 \end{equation}
where $A^j$ are arbitrary constants, contains finite number of
operators, $1 \leq j \leq l$, the group is called
\textit{finite-dimensional} (or simply finite) with the dimension
$l$; otherwise, for unlimited $j$ or in the case that coordinates
$\xi^{i}$, $\eta^{\alpha}$ depend upon arbitrary functions of
group variables, the group is called \textit{infinite}. \par

The use of the infinitesimal criterion (\ref{deteq}) for
calculating the symmetry groups makes the whole procedure
algorithmic and can be carried out not only ``by hand" but using
the symbolic packages of the computer algebra (see, e.g., Vol.3 in
\cite{Sprav}) as well. In modern group analysis, different
modifications of the classical Lie scheme are in use (see
\cite{Oves,Olver,Sprav} and references therein).

The generator (\ref{oper}) of the group $\cal{G}$ is equivalent to
the canonical Lie--B\"acklund operator
\begin{equation}
\label{lboper}
 Y=\kappa^\alpha \partial_{u^{\alpha}}\,, \quad
 \kappa^\alpha \equiv \eta^{\alpha} - \xi^{i} u_{i}^{\alpha}  \,,
\end{equation}
that is known as a \textit{canonical representation} of $X$ and
plays an essential role in RGS constructing.

However, the group defined by the generators (\ref{oper}) and
(\ref{lboper}) cannot yet be referred to as a renormgroup, as it
is not related to a partial BVP solution of interest. \par
\smallskip

\textbf{III.}  To obtain RGS, the restriction of the group
$\cal{G}$ on particular BVP solution should be made which forms
the \textbf{third} step. Mathematically, this procedure appears as
checking the vanishing condition for the linear combination of
coordinates $\kappa_j^{\alpha}$ of the canonical operator
equivalent to (\ref{gen-oper}) on a particular approximate (or
exact) BVP solution $U^{\alpha}(z)$
 \begin{equation} \label{restrict}
{\left\{\ \sum\limits_{j} A^j \kappa^{\alpha}_j   \equiv
 \sum\limits_{j} A^j \left(\eta^{\alpha}_j-\xi^i_j
 u_i^{\alpha}\right)\ \right\}}_{
 \Big\vert \displaystyle{u^{\alpha}=U^{\alpha}(z)}}=0 \,.
 \end{equation}
Evaluating (\ref{restrict}) on a particular BVP
solution $U^{\alpha}(z)$ transforms the system of DEs for group
invariants into algebraic relations\footnote{Similar relations
were discussed in \cite{Oves}, Chapter 8, when constructing
invariant solutions for the Cauchy problem for a quasi-linear
system of first order PDEs.}. Firstly, it gives relations between
$A^j$ thus ``combining" different coordinates of group generators
$X_j$ admitted by the $\cal{RM}$ (\ref{rgm}). Secondly, it
eliminates (partially or entirely) the arbitrariness that may
appear in coordinates $\xi^{i}$, $\eta^{\alpha}$ in the case of an
infinite group $\cal{G}$. In terms of the ``classic" QFT RG
terminology, where it exists only one operator $X$ of
(\ref{bereg2})--type (i.e., all $A_j$ except one are equal to
zero), the procedure of group restriction on a particular BVP
solution $\bar{g}_{appr}$ eliminates arbitrariness in the form of
$\beta (g)$--function. \par

While the general form of the condition given by
Eq.(\ref{restrict}) is the same for any BVP solution, the way of
realization of the restriction procedure in every particular case
employs a particular perturbation approximation (PA) for the
concrete BVP. \par

Generally, the restriction procedure reduces the dimension of
$\cal{G}$. It also ``fits" boundary conditions into the operator
(\ref{gen-oper}) by a special choice of coefficients $A_j$ and/or
by choosing the particular form of arbitrary functions in
coordinates $\xi^{i}$, $\eta^{\alpha}$. Hence, the general element
(\ref{gen-oper}) of the group $\cal{G}$ after the fulfillment of a
restriction procedure is expressed as a linear combination of new
generators $R_{i}$ with the coordinates $\tilde\xi^{i}$,
$\tilde\eta^{\alpha}$,
 \begin{equation} \label{rgo}
 X\ \Rightarrow \ R = \sum\limits_{j} B^j R_j \,, \quad
 R_{j} = \tilde\xi^{i}_{j}\partial_{x^i} + \tilde\eta^{\alpha}_{j}
         \partial_{u^{\alpha}}\,,
 \end{equation}
where $B^j$ are arbitrary constants.\par

The set of  RGS generators $R_i$ each containing the desired BVP
solution in its invariant manifold, define a group of
transformations that we also refer to as \textit{renormgroup}.
Therefore, here we \textsf{extend the notion of renormgroup and RG
symmetry} and the direct analogy with the ``Bogoliubov RG" is
preserved only for one-parameter group of point transformations.
\par
\medskip

\textbf{IV.}~ The prescribed three steps entirely define the
regular algorithm of RGS construction but do not touch on how a
BVP solution is found. Hence, one more important, the
\textbf{fourth}, step should be added. It consists in using RGS
generators to find analytical expressions for the new,
``improved", solution of the BVP. \par

Mathematically, this step makes use of RG=FS \textit{invariance
conditions} that are given by a combined system of (\ref{rgm}) and
the vanishing condition for the linear combination of coordinates
$\tilde\kappa^{\alpha}_{j}$ of the canonical operator equivalent
to (\ref{rgo}),
 \begin{equation} \label{fsinv}
 \sum\limits_{j} R^j \tilde\kappa^{\alpha}_j   \equiv
 \sum\limits_{j} B^j \left(\tilde\eta^{\alpha}_j-\tilde\xi^i_j
 u_i^{\alpha}\right)\ =0 \,.
 \end{equation}
One can see that conditions (\ref{fsinv}) are akin to
(\ref{restrict}). However, in contrast with the previous step, the
differential variables $u$ in (\ref{fsinv}) should not be replaced
by an approximate expression for the BVP solution $U(z)$, but
should be treated as usual dependent variables. \par

For the \textit{one-parameter Lie point renormgroup}, RG
invariance conditions lead to the \textit{first order PDE} that
gives rise to the so-called \textit{group invariants} (like
invariant couplings in QFT) which arise as solutions of associated
characteristic equations.  \par

A general solution of the BVP is now expressed in terms of these
invariants. On the one hand, this is in direct analogy with the
structure of RG invariant solutions in QFT -- compare with
Eqs.(\ref{s-rg2}) and (\ref{genRGsol}).\par

On the other hand, it reminds the so-called $\Pi$--theorem from
the theory of dimensional analysis and similitude (see, Section 19
in \cite{Oves}, Section 6 of Chapter 1 in \cite{Sedov} and
historical comment to Section 43 in \cite{Birkhoff60}) directly
related to power self-similarity, discussed above in Section
\ref{ss2.4}. \par

However, as we shall see later, in the general case of arbitrary
RGS the group invariance condition obtained for BVP is not
necessarily characteristic equations for the Lie point group
operator. They may appear in a more complicated form, e.g., as a
combination of PDE and higher order ODE (see Section \ref{ss4.2}).
Nevertheless, the general idea of finding solution of the BVP as
RG invariant solutions remains valid.

\subsection{Examples of solution improving \label{ss3.2}}

We present now a few examples of the RGS construction with further
use of the symmetry for ``improving" an approximate solution.

\subsubsection{\small\it Modified Burgers equation \label{sssBurgers}}

  As the first example, we take the initial value problem for the modified
Burgers equation
 \begin{equation} \label{bureq}
 u_t-au_x^2-{\nu}{u}_{xx}=0\,, \quad  u(0,x)=f(x) \,. \end{equation}
It is connected with the heat equation
\begin{equation} \label{heat}
 \tilde u_t= \nu \tilde u_{xx}  \end{equation}
by transformation $\tilde{u}=\exp(au/\nu)$ and has an exact
solution.  Due to this, while using RGS to find a solution, one
can check the validity of our approach. The RGS constructing for
(\ref{bureq}) is an apt illustration of the general scheme, shown
in Figure 1 which may be helpful in understanding other examples
of the general algorithm implementation. We review here in short
the procedure and results of paper Ref.\cite{Burgers}.
\par

The RG-manifold $\cal{RM}$ (step {\textbf{I}}) is given by
Eq.(\ref{bureq}) with the parameters of nonlinearity $a$ and
dissipation $\nu$ included in the list of independent variables.
The Lie calculational algorithm applied to $\cal{RM}$ gives, for
the admitted group $\cal G$ (step \textbf{II}),
 nine independent terms in the general expression for the group generator
\begin{equation}  \label{burgroup}
 X= \sum \limits_{i=1}^{8} A^{i}(a,\nu) X_i + \alpha (t,x,a,{\nu})
 e^{ -au/\nu}  \, \partial_u \, , \end{equation}
 \[
 \begin{array}{l}
 \displaystyle{
  X_1=4{\nu}t^2{\partial}_{t}+4{\nu}tx{\partial}_{x}
      -({\nu}/a)(x^2+2{\nu}t){\partial}_{u}\, , \quad
  X_2 =2t{\partial}_{t}+x{\partial}_{x}\, , } \\
 \mbox{}\\
 \displaystyle{
   X_3=(1/{\nu}){\partial}_{t} \, , \quad
   X_4 =2{\nu}t{\partial}_{x}-({\nu}/a)x{\partial}_{u}\, , \quad
   X_5={\partial}_{x}\, , \quad
   X_6=-({\nu}/a){\partial}_{u} \, ,} \\
 \mbox{}\\
 \displaystyle{
   X_7 =a{\partial}_{a}+\left[ ({\nu}/a)
        - u \right] {\partial}_{u}\,, \quad
   X_8 =2{\nu}{\partial}_{\nu}+x{\partial}_{x}
       +2 \left[ u - ({\nu}/a) \right] {\partial}_{u} \, . }\\
 \end{array}
 \]

Here, $A^i(a, \nu)$ are arbitrary functions of their arguments and
$\alpha(t,x,a,{\nu})$ is an arbitrary function of four variables,
satisfying the heat equation (\ref{heat}).

  A set of operators $X_i$ form an eight-dimensional Lie algebra
$L_8$. The first six generators relate to the well-known
symmetries of the modified (potential) Burgers equation (see,
e.g., Vol.1, p.183 in \cite{Sprav}). They describe projective
transformation in the $(t,x)-$plane ($ X_1$), dilatations in the
same plane ($ X_2$), translations along the $t-,\,x-$ and $u-$axes
($ X_3$, $X_5$ and $X_6$) and Galilean transformations ($X_4$).
The last two generators $X_7$ and $ X_8$ relate to dilatations of
parameters $a$ and $\nu$ now involved in group transformations.

  The procedure of restriction (step {\textbf{III}}) of the group
(\ref{burgroup}) admitted by $\cal{RM}$ (\ref{bureq}) implies the
check of the invariance condition (\ref{restrict}) on a
particular BVP solution $u=U(t,x,a,\nu)$
 \begin{equation}  \label{burfscond}
 {\left\{ {\eta}_{\infty} + \sum\limits_{i=1}^{8}  A^i(a,\nu)
 \, \kappa_{i} \right\}}_{\Big\vert {\displaystyle{u=U(t,\, x,\, a,\,\nu)}}}=0 \,, \quad
 \kappa_i \equiv {\eta}_{i}-{\xi}^{1}_{i} u_t
 - {\xi}^{2}_{i} u_x - {\xi}^{3}_{i} u_a - {\xi}^{4}_{i} u_{\nu} \,.
 \end{equation}
This formula expresses the coordinate $\alpha$ of the last term in
(\ref{burgroup}) via the remaining coordinates of eight generators
$X_i$ for arbitrary $t$, and hence for $t=0$, when $\
U(0,x,a,\nu)=f(x)$.  As a result, we obtain the ``initial" value
$\alpha(0,x,a,\nu)$ and then, using the standard representation
for the solution to the linear parabolic equation (\ref{heat}),
the value of $\alpha$ at arbitrary $t\neq 0$
  \begin{equation}   \label{ideal}
 \alpha (t,x,a,\nu) = - \sum\limits_{i=1}^{8} A^i(a,\nu)
 <{\overline {\kappa}_i (x,a,\nu)} > \, .   \end{equation}
Here, $\overline {\kappa}_i (x,a,\nu)$ denote ``partial" canonical
coordinates $\kappa_i$ taken at $t=0$ and $u=f(x)$. Symbol $<F>$
designates the convolution of a function $F$ with the fundamental
solution of (\ref{heat}), multiplied by the exponential function
of $f$ entering into the boundary condition $$
 <F(x,t,a,\nu)> \equiv \frac{1}{\sqrt{4\pi \nu t}}
 \int\limits_{-\infty}^{\infty}\,dy\, F(y,t,a,\nu) \exp
 \left(-\frac{(x-y)^2}{4 \nu t} + \frac{a f(y)}{\nu}\right) \,.   $$
Substitution (\ref{ideal}) in the general expression
(\ref{burgroup}) gives the desired RG generators
\[
 R_{i}=X_{i}+ \varrho_i e^{-au/\nu}\, \partial_u\,,
\]
\[
\begin{array}{l}
\displaystyle{\varrho_1 =\frac{\nu}{a}<x^2> \,, \quad
 \varrho_2 =<xf_x> \,, \quad \varrho_3=\frac{1}{\nu}<af_x^2+\nu f_{xx}>\,,
 \quad \varrho_4 = \frac{\nu}{a}<x> \,,}\\   \mbox{}\\
\displaystyle{\varrho_5 =<f_x> \,, \quad \varrho_6 =
\frac{\nu}{a}<1> \,, \quad \varrho_7 = <f -\frac{\nu}{a}> \,,
\quad
 \varrho_8 =\, <xf_x -2f+2\,\frac{\nu}{a} > \,.}  \end{array}
\]
Operators $R_i$ form an \textit{ eight-dimensional RG algebra} $RL_8$ that
has the same tensor of structural constants as $L_8$, i.e. $RL_8$ and $L_8$
are isomorphic. Hence, the procedure of the group
restriction eliminates the arbitrariness presented by the function
$\alpha$ and ``fits" the boundary conditions into RG generators by
means of $\varrho_i$.
\par

It can be verified that the exact
solution of the initial-value problem (\ref{bureq})
\begin{equation} \label{bursol}
u(t,x;a,\nu)
=\frac{\nu}{a}\,\ln <1>\equiv \frac{\nu}{a}\,\ln\left\{\frac{1}
{\sqrt{4\pi\nu t}} \int\limits_{-\infty}^{\infty}\,dy\, \exp
 \left(-\frac{(x-y)^2}{4 \nu t} + \frac{a f(y)}{\nu}\right)  \right\}
\end{equation}
is the invariant manifold for any of the above RGS operators. And
vice versa, (\ref{bursol}) can be reconstructed from an
approximate solution with the help of any of the RGS operators or
their linear combination. For example, two such operators, $\nu
R_3 \equiv R_t$ and $(1/a)(R_6 + R_7)\equiv R_a$ were used in
\cite{Burgers} to reconstruct the exact solution from perturbative
(in time and in nonlinearity parameter $a$) solutions. Below, we
describe this procedure (step {\textbf{IV}}) using the operator
$R_a$,
\begin{equation}  \label{burrg-a}
R_a= \partial_a + (1/a)\left(-u+e^{-au/\nu} <f(x)> \right)
\partial_u \,.
 \end{equation}
It is evident that $t, x$ and $\nu$ are invariants of group
transformations with (\ref{burrg-a}), whilst finite RG
transformations of the two remaining variables, $a$ and $u$, are
obtained by solving the Lie equations for (\ref{burrg-a}), with
$\ell$ the group parameter
  \begin{equation}   \label{burlieeq}
 \frac{d a^\prime}{d \ell}=1 \, ,\quad a^\prime\vert\, {}_{\ell=0}= a \, ;
  \quad
 \frac{d u^\prime}{d \ell}=\alpha(t,x,a^\prime, \nu)\,
  e^{-a^\prime u^\prime/\nu} - \frac{u^\prime}{a^\prime} \, ,\quad
  u^\prime \vert \, {}_{\ell =0} = u \, .    \end{equation}
Combining these equations yields one more invariant
${\cal{J}}=e^{au/\nu}- <1>$ for the RGS generator (\ref{burrg-a}).
Solution of (\ref{burlieeq}) along with (\ref{ideal}) gives the
formulae for finite RG transformations of the group variables $\{
t,x,a,\nu,u \}$
\begin{equation} \label{burtr}
\begin{array}{l}
 \displaystyle{ t^\prime=t\,,\ x^\prime=x \,,\ \nu^\prime=\nu  \,,
  \ a^\prime  =  a + \ell  \,, \
  u^\prime=\frac{\nu}{a+\ell}\ln\left( e^{au/\nu}
  + <e^{\ell f(x)/\nu}-1>\right)\,.}
\end{array} \end{equation}
Choosing the value $a$ equal to zero, which is a starting point of
PA in $a$, we get $a^\prime=\ell$. Then after excluding $t,\,
x,\,\nu$ and $\ell$ from the expression for $u^{\prime}$
(\ref{burtr}) and omitting accents over $t^\prime,\, x^\prime\,,
\nu^\prime\,, u^\prime $ and $a^\prime$ the desired BVP solution
(\ref{bursol}) is obtained. It also follows directly from
$\cal{J}$ in view of the initial condition
${\cal{J}}\mid_{a=0}=0$.

A similar procedure can be fulfilled for the other RG operator,
\[
R_t =\partial_t+ e^{-au/\nu} <af_x^2+\nu f_{xx}>\partial_u\,,
\]
which is consistent with the PA in time $t$. Although invariants
for $R_t$ and finite RG transformations
differ from that for (\ref{burrg-a}), the final result, i.e., the
exact solution of BVP (\ref{bureq}) given by (\ref{bursol}), is
the same. This possibility is the distinct demonstration of the
multi-dimensional RGS to reconstruct the \textit{unique} BVP
solution from \textit{different} PA: either in
parameter $a$ or in $t$ (though we used only two one--dimensional
subalgebras here\footnote{This can be considered as a construction
parallel to the one used in Ref.\cite{cris98}.}).

\subsubsection{\small\it BVP for ODE: simple example \label{sssGolden}}

Quite recently, the QFT renormalization group ideology has been
applied, a bit straightforward, in mathematical physics for
asymptotic analysis of solutions to DEs \cite{gold96,Bric94} and
in constructing an envelope of the family of solutions
\cite{Kunih}. \par

 Our second methodological example with linear ODE is presented here in
order to illustrate the difference between our approach and the
``perturbative RG theory" devised in \cite{gold96} for a global
analysis of BVP solutions in mathematical physics. \par

Consider a linear second order ODE for $y(t)$ with the initial
conditions at $t=\tau$,
\begin{equation} \label{NG}
y_{tt} + y_{t} + \varepsilon y = 0\,,      \quad y(\tau) = \tilde
{u}\,,
    \quad y_t(\tau)= \tilde {w}\,,  \end{equation}
which has the exact solution:
 \begin{equation}  \label{exact-sol}
      y= C_+ e^{-\gamma_+ (t-\tau)} + C_- e^{-\gamma_- (t- \tau )} \,,
       \  \gamma_{\pm} =\frac{1\pm K}{2} \,,
       \  K = \sqrt{1-4\varepsilon}\,, \
   C_{\pm}=\mp\frac{\tilde{w}+\gamma_{\mp}\tilde{u}}{K}\,.
 \end{equation} \par

Provided that the parameter $\varepsilon$ is small, the solutions
to Eq.(\ref{NG}) has been treated in \cite{gold96} with the aim to
demonstrate effectiveness of the ``perturbative RG theory" for an
asymptotic analysis of a solution behaviour. The main goal of this
treatment was to improve a perturbative expansion in powers of
$\varepsilon$  with secular terms $\propto\varepsilon(t-\tau)$ and
obtain\footnote{ The algorithm used in \cite{gold96} for improving
PA solutions with secular terms involves \textit{a)} an
introduction of some additional parameters in solutions,
\textit{b)} a special choice of these parameters that eliminates
secular divergencies, and \textit{c)} imposing a condition of
independence of a solution upon the way of introducing these
parameters. In some cases, this algorithm, directly borrowed from
QFT RG-method, gives an exact solution. However, the question of
correspondence of this construction to a transformation group of a
solution of BEs remains open.} a uniformly valid asymptotic of a
solution
\begin{equation}  \label{Gold-sol}
 y=  c_{+} e^{ ( - 1 + \varepsilon (1 + \varepsilon)) \, ( t - \tau ) }
  + c_{-} e^{-\varepsilon(1+\varepsilon)\, (t-\tau)} + O(\varepsilon^2) \,,
\end{equation}
\[
 c_{+} \approx-
     \left((1+2\varepsilon)\tilde{w}+\varepsilon\tilde{u}\right) \,,
 \quad
 c_{-} \approx
     \left((1+2\varepsilon)\tilde{w}+(1+\varepsilon)\tilde{u}\right)\,,
\]
which is accurate for small values $\varepsilon \ll 1$ but for
arbitrary values of the product $\varepsilon(t -\tau)$.
\smallskip

We are going to show that the use of our regular RG algorithm
enables one to improve a PA solution (either in powers of
$\varepsilon$ or in $t-\tau$) up to the \textsf{exact} BVP
solution (\ref{exact-sol}).
\par

Rewriting (\ref{NG}) in the form of the system of two first order
ODEs for functions $u\equiv y $ and $w\equiv y_t $,
 \begin{equation} \label{ngo}
 u_{t} = w\,,  \quad w_{t} = - \varepsilon w - u \,,
 \end{equation}
we construct $\cal{RM}$ (step \textbf{I}) using the
\textit{invariant embedding method} (this approach has first been
realized in \cite{KKP_RG-91}). Then, $\cal{RM}$ is presented as a
joint system of BEs (\ref{ngo}) and embedding equations
\[
    u_{\tau} -(\varepsilon {\tilde w} + {\tilde u})u_{\tilde w}
               - {\tilde w} u_{{\tilde u}} = 0 \,,
\quad  w_{\tau} -(\varepsilon {\tilde w} + {\tilde u})w_{\tilde w}
               - {\tilde w} w_{{\tilde u}} = 0 \,,
\]
treated in the extended space of group variables which include the parameters
 $\tau, \ \tilde{w},\ \tilde{u}$ of boundary conditions in addition to $t$
and dependent variables $u,\ w $.

 Omitting tedious calculations related to the following two steps
(steps \textbf{II} and \textbf{III} ), we present here two examples of
resulting RGS generators
\begin{equation}   \label{embrg1}
\begin{array}{l}
 \displaystyle{   R_{\tau} = \partial_{\tau}
     - ( {\tilde w} + \varepsilon {\tilde u} )\partial_{\tilde w}
     + {\tilde w} \partial_{\tilde u}}  \,,\\      \mbox{}\\
 \displaystyle{  R_{\varepsilon}= \partial_{\varepsilon}
  - \left(\frac{1}{\mu^2} (2{w} + {u}) (1-t/2)
  + (t/2) {u} \right) \partial_{w}
  + \frac{t}{\mu^2}\left( 2{w} + {u} \right) \partial_{u}}\\  \mbox{}\\
 \displaystyle{\hphantom{R_5= \partial_{\varepsilon}}
  - \left( \frac{1}{\mu^2} (2\tilde{w} + \tilde{u}) (1-\tau/2)
  + (\tau/2) \tilde{u} \right) \partial_{\tilde w}
 +\frac{\tau}{\mu^2}\left(2\tilde{w}+\tilde{u}\right)\partial_{\tilde u}\,,}
 \end{array}
\end{equation}
that involve the initial values $\tilde w$, $\tilde u$ and initial
point $\tau$ in RG transformations. In addition, $R_{\varepsilon}$
transforms the parameter $\varepsilon$. \par

Now, the procedure of constructing the BVP solution
(\ref{exact-sol}) (step \textbf{IV}) is similar to that used in
the previous Section \ref{sssBurgers} and employ finite
transformations that are defined by the Lie equations for the
operators (\ref{embrg1}). For $R_{\tau}$ functions $u, \ w$ and
the parameter $\varepsilon$ are group invariants, while the
translations of $\tau $ and the corresponding transformations of
$\tilde{u}, \ \tilde{w}$ restores the exact solution
(\ref{exact-sol}) from the PA in powers of $t-\tau$ (note that the
parameter $\varepsilon$ is not necessarily small in this PA!).
\par

For $R_{\varepsilon}$ the difference $t-\tau$ is  group invariant,
whilst the transformation of $\varepsilon$ and related
transformations of $u,w,\tilde{u}, \tilde{w}$ restore the exact
solution (\ref{exact-sol}) from the PA (discussed in
\cite{gold96}) in powers of $\varepsilon$. Hence, as in the
previous Section \ref{sssBurgers}, both the RGS generators
(\ref{embrg1}) reconstruct the unique BVP solution from different
PAs.

\section{RG in Nonlinear optics \label{s4}}
\subsection{Formulation of a problem \label{ss4.1}}

As a problem of real physical interest, take BVP that
describes self-focusing of a high-power light beam. While the
problem plays an important role in nonlinear electrodynamics since
60s, the detailed quantitative understanding of
self-focusing is still missing \cite{PRA}, and there is no method
which allows to find an analytic solution to the corresponding
equations with arbitrary boundary conditions.\par

Here, we demonstrate the great potential of the RGS approach in
constructing analytic solutions of BVP equations with arbitrary
boundary conditions. The RGS method allows to consider different
types of BEs for self-focusing process which include plane and
cylindrical beam geometry, nonlinear refraction and diffraction.
The merit of the RGS method is that it describes BVP solutions
with one-- or two--dimensional singularities in the entire range
of variables from the boundary up to the singularity point.
\par

Let us start with BVP for the system of two DEs
\begin{equation} \label{basic}
 v_{z}+v v_{x}-\alpha n_{x} = 0 \,, \quad
 n_{z}+n v_{x}+v n_{x}+(\nu-1)\left( nv/x \right) = 0\,,
 \end{equation}
 \begin{equation} \label{boundary}
 v(0,x) =  0 \,, \qquad  n(0,x)=N(x) \,,
 \end{equation}
which are used in nonlinear optics of self-focusing wave beams
when diffraction is negligible. \par

We study spatial evolution of the derivative of the beam eikonal
$v$ and the beam intensity $n$ in the direction inwards the medium
$z$ and in the transverse direction $x$. The term proportional to
$\alpha $ is related to nonlinear refraction effects; $\nu=1$ and
$\nu=2$ refers to the plane and cylindrical beam geometry,
respectively. Boundary conditions (\ref{boundary}) correspond to
the plane front of the beam and the arbitrary transverse intensity
distribution.  \par

\subsection{Plane geometry \label{ss4.2}}

In the plane beam geometry (at $\nu=1$) Eqs.(\ref{basic}) can be reduced
to the system of BEs
 \begin{equation} \label{nlopeq2}
 \tau_w-n\chi_n=0\,,\quad\chi_w +\alpha \tau_n = 0 \,,
 \end{equation}
for functions $\tau= n z$ and $\chi=x-v z$ of $w=v/\alpha$ and $n$
arguments, with boundary conditions
 \begin{equation} \label{nlopcond2}
 \tau(0,n)=0 \,, \quad \chi(0,n)=H(n)\,,
 \end{equation}
where $H(n)$ is the inverse to $N(x)$. Here, the procedure of RGS
constructing makes use of the {\it Lie--B\"acklund symmetry} and
is described as follows \cite{KPSh-JMP98}. The manifold $\cal{RM}$
(step {\textbf{I}}) is defined by Eqs.(\ref{nlopeq2}) treated in
the extended space that include dependent and independent
variables $\tau ,\ \chi,\ w,\ n$ and derivatives of $\tau$ and
$\chi$  with respect to $n$ of an arbitrary high order. The
admitted symmetry group $\cal G$ (step {\textbf{II}}) is
represented by the canonical Lie-B\"acklund operator
\begin{equation}
 X = f\partial_{\tau}+g\partial_{\chi}\,,\label{oprgnl1}
\end{equation}
with the coordinates $f$ and $g$ that are linear combinations of $\tau$ and
$\chi$ and their derivatives $\partial^i\tau/\partial n^i$ and $\partial^i
\chi / \partial n^i \,,~i \geq 1$ with the coefficients depending on $w$ and
$n$.\par

The restriction of the group admitted by $\cal{RM}$
(\ref{nlopeq2}) (step {\textbf{III}})) implies the check of the
invariance condition (\ref{restrict}) that yields two relations
 \begin{equation}       \label{fssnl1}
 f=0\,, \quad g=0 \,.
 \end{equation}
These relations should be valid on a particular solution of
BVP with the boundary data (\ref{nlopcond2}). For example, choosing
the so-called ``soliton" profile, $N(x)=\cosh^{-2}(x)$, i.e.,
$H(n)=\mathrm{Arccosh} \left( 1/ \sqrt{ {\,} n }
\right)$, we have
 \begin{equation}   \label{rgs1-soliton}
 \begin{array}{ll}
 \displaystyle{ f = } &  \displaystyle{  2n(1-n)\tau_{nn}- n\tau_{n}
         - 2nw(\chi_{n} + n \chi_{nn} )
         + \left(\alpha w^2/\, 2 \right) n \tau_{nn} \,, }  \\   \mbox{}\\
 \displaystyle{ g= } &   \displaystyle{  2n(1-n)\chi_{nn} +(2-3n)\chi_{n} +
                \alpha w \left( 2n\tau_{nn} + \tau_{n} \right)
         + \left( \alpha w^2 /\, 2 \right) \left( n \chi_{nn} + \chi_{n} \right) \,. }
 \end{array}
 \end{equation}
Dependence on $\tau_{nn}$ and $\chi_{nn}$ indicates that here RGS
is \textit{the second-order Lie-B\"acklund symmetry}. In order to
find a particular solution of a BVP (step {\textbf{IV}}), one
should solve the joint system of BEs (\ref{nlopeq2}) and
second-order ODEs that follow from the RG=FS invariance conditions
(\ref{fssnl1}) and (\ref{rgs1-soliton}). The resulting expressions
\cite{K-TMF97} -- the well-known Khokhlov solutions\footnote{In
Ref.\cite{Akhmanov}, where this solution was first obtained, it
did not result from a regular procedure.}
 \begin{equation}   \label{soliton}
 v=-2\alpha nz \tanh (x-vz)\,, \quad  \alpha n^2 z^2= n\cosh^2(x-vz)-1\,,
 \end{equation}
describe the process of self-focusing of a soliton beam: the
sharpening of the beam intensity profile with the increase of $z$
is accompanied by the intensity growth on the beam axis. The
solution (\ref{soliton}) is valid up to the singularity point
where the derivatives $v_x$ and $n_x$ tend to infinity whilst the beam
intensity $n$ remains finite
\begin{equation}  \label{sing-soliton}
 z_{sing}^{sol}=1/2\sqrt{\, \alpha} \,, \quad  n_{sing}^{sol} = 2\,.
\end{equation}
\indent Here, the Lie-B\"acklund RGS enables one to reconstruct the
BVP solution and describe the solution singularity for the light
beam with the soliton initial intensity profile. One more example
of an {\it exact} BVP solution obtained with the help of  Lie-B\"acklund
RGS (with the initial beam profile in the form of a "smoothed"
step) can be found in \cite{K-TMF97}.
 \smallskip

For arbitrary boundary data, it turns to be impossible to fulfill
the condition (\ref{fssnl1}) with the help of the Lie-B\"acklund
symmetries of any finite order, and one is forced to use a
different algorithm \cite{KPSh-JMP98,K-TMF97} of RGS constructing,
based on the approximate group methods. Here, (step {\textbf{I}}))
$\cal{RM}$ is given by BEs (\ref{nlopeq2}) with a small parameter
$\alpha$, and coordinates of the group generator (\ref{oprgnl1})
(and, hence, coordinates of the RGS operators) appear as infinite
series in powers of $\alpha$
 \begin{equation} \label{coord1}
   f=\sum\limits_{i=0}^{\infty} \alpha^i f^i \,; \quad
   g=\sum\limits_{i=0}^{\infty} \alpha^i g^i\,.     \end{equation}
The procedure of finding the coefficients $f^i,~g^i$ (step
{\textbf{II}})) leads to the system of recurrent relations that
express higher-order coefficients $f^{i+1}$, $g^{i+1}$ in terms of
previous ones $f^{i}$, $g^{i}$. It means that once the zero-order
terms are specified, the other terms are reconstructed by the
recurrent relations.  \par

The coefficients $f^i$ and $g^i$ contain an arbitrary function of
$n, \chi_{[s]}$ and ${\tau}_{[s]}-w(s\chi_{[s]}+n\chi_{[s+1]})$
where subscript $[s]$ denotes the partial derivative of the order
$s$ with respect to $n$. This arbitrariness is eliminated by the
procedure of group restriction (step \textbf{III})), i.e., by
imposing the invariance condition (\ref{fssnl1}). For particular
forms of $f^0$ and $g^0$, that is for partial boundary conditions
(\ref{nlopcond2}), \textit{infinite} series are \textit{truncated}
automatically, and we arrive at the \textit{exact RGS}.
\par

One example of this kind is given by Eqs.(\ref{rgs1-soliton}) that have
a binomial structure $f=f^0+\alpha f^1$, $g=g^0+\alpha g^1 $. If
we neglect the higher-order terms in the case of arbitrary boundary
conditions (when series (\ref{coord1}) are not truncated
automatically), then we get an \textit{approximate RGS} which
produces an approximate solution to the BVP. As an example, we
give here two sets of expressions for the coordinates $f^i$ and $g^i$
for the Gaussian initial profile with $N(x)=\exp(-x^2)$, i.e.,
 $H(n)=\left( \ln (1/n)\right)^{1/2}$, which define approximate RGS
\[
 \begin{array}{ll}
 \displaystyle{\mathbf{ a)}} &
 \displaystyle{  \ f^0 = 1+ 2n\chi \chi_n \,,  \quad g^0=0\,,
 \quad f^1 = -2 \tau \tau_n +\frac{\tau^2}{n} \,, \quad g^1= -2
 \left(\tau \chi_n + \chi \tau_n  \right) \,,} \\     \mbox{}\\
 \displaystyle{ \mathbf{b)} } &
 \displaystyle{  \ f^0 =  2n(\tau \chi_n
 + \tau_n\chi) \,,\ \ g^0 = 1 + 2n\chi \chi_n\,,} \ \
 \displaystyle{ f^1 = 2 \chi \tau_{\alpha} \,,\ \ g^1= 2 \left(\chi
 \chi_{\alpha} - \tau \tau_n \right) \,.}
 \end{array}
\]

    Here, linear dependencies of $f$ and $g$ upon first-order
derivatives indicate that RGS is equivalent to Lie
point symmetry. The peculiarity of the case $\mathbf{b}$ is a
dependence of $f$ and $g$ not only on derivatives with respect to
$n$ but also with respect to $\alpha$: it means that the parameter
$\alpha$ is also involved in group transformations. In the
non-canonical representation (\ref{oper}), the RGS generator in
this case has the form
 \begin{equation}   \label{rg-gsa}
 R_{Guass1}=  2\tau \partial_{w} + 2n\chi \partial_{n}+
        2\alpha\chi\partial_{\alpha} - \partial_{\chi} \,. \end{equation}

The last step \textbf{IV} is performed in a usual way by solving the joint
system of BEs (\ref{nlopeq2}) and equations that follow from the RG=FS
invariance condition (\ref{fssnl1}), or else, using invariants of associated
characteristic equations for RG operator provided that RGS is a
Lie point symmetry. We give here the solution that follows from
RGS (\ref{rg-gsa}),
 \begin{equation} x^2=\left( 1 - 2 \alpha nz^2 \right)^2 \ln
     \frac{1}{n(1- \alpha nz^2)} \,, \quad
 v = - \frac{ 2x\, \alpha nz } {1-2 \alpha nz^2} \,.
 \label{res2_gauss}
 \end{equation}
These expressions describe a self-focusing Gaussian beam (the plot
$n(x)$ for this solution is presented at the end of the section on
Figure 2), that is qualitatively very similar to the spatial
evolution of the soliton beam (\ref{soliton}). Moreover, the
singularity position and the value of maximum beam intensity at
this point coincide with analogous values (\ref{sing-soliton}) for
the soliton beam. Although formulae (\ref{res2_gauss}) correspond
to an approximate BVP solution, they exactly describe the
behaviour of $n$ on the beam axis at $x=0$. To estimate the
reliability of result (\ref{res2_gauss}) in the off-axis region,
we compared it with another approximate BVP solution which arises
from the approximate RGS in the case $\mathbf{a}$. These
approximations agree very well (details are presented in
\cite{K-TMF97}), thus proving the accuracy\footnote{One more
evidence is provided by the comparison of approximate and exact
BVP solution for the soliton beam performed in \cite{K-TMF97}.} of
the RG approach.
\par

\subsection{Cylindrical geometry \label{ss4.3}}

In the above discussion we dealt with the plane beam geometry and took
into account only effects of nonlinear beam refraction, neglecting
diffraction. The flexibility of RGS algorithm allows one to apply it in a
similar way to a more complicated model as compared to (\ref{nlopeq2}),
e.g., for the cylindrical beam geometry, $\nu=2$. Omitting technical details,
we present the RGS generator for the cylindrical parabolic beam with $N=1-x^2$
\[
 R_{par} = \left( 1-2\alpha z^2 \right) \partial_{z}
    -2\alpha zx \partial_{x}
    - 2\alpha \left( x-vz \right) \partial_{v}
    + 4\alpha nz \partial_{n} \,.
\]

The BVP solution is expressed in terms of group invariants for this
generator:
\begin{equation}    \label{invariants}
{\cal{J}}_{1}=\frac{x^2}{\varrho} \,; \quad {\cal{J}}_{2}=n
\varrho \,; \quad
 {\cal{J}}_{3} = 2 \alpha x^2 - v^2 \varrho +  \frac{x v}{2} \varrho_z \,;
  \quad  \varrho =  \left( 1-2\alpha z^2 \right) \,.  \end{equation}

 The explicit form of dependencies of ${\cal{J}}_2=1-{\cal J}_1^2 \,,~
{\cal J}_3=2\alpha{\cal J}_1\,$ upon ${\cal J}_1$ follows from the
boundary conditions (\ref{boundary}). They lead to the well-known
solution \cite{Akhmanov}
\begin{equation}  \label{parabolic}
 v=(x/(2\varrho))\varrho_z\,, \qquad n=(1/\varrho)(1-(x^2/\varrho))
\end{equation}
that describes the convergence of the beam to the singularity
point $z_{sing}^{par}=1/\sqrt{2\alpha}$ where $\varrho=0$ and $
n\to\infty$. The solution singularity is two-dimensional here: the
infinite growth of beam intensity in the vicinity of the
singularity $z \to z_{sing}^{par}$ is accompanied by the infinite
growth of the derivative $v_x$ and collapsing of the beam size in
the transverse direction.

The RGS algorithm based on approximate group methods can also be
applied in the case when besides nonlinear refraction also
diffraction effects are taken into account. Then, the first
equation in (\ref{basic}) should be modified by adding the
diffraction term
 \[
 -\beta\partial_x\left\{\left(x^{1-\nu}/\sqrt{\, n\,}\right)
  \partial_x \left(x^{\nu-1}\partial_x \sqrt{\, n\,} \right)\right\}\,.
  \]

 Standard calculations done in compliance with a general scheme for thus
modified $\cal{RM}$ (for details see \cite{K-TMF99}) give the RGS
generator for the cylindrical beam geometry $(\nu=2)$
\begin{equation}   \label{rgsym}
 R_{Gauss2} = \left( 1 + z^2 S_{\chi\chi} \right)\partial_{z}
   + \left( z S_{\chi} + v z^2 S_{\chi\chi} \right) \partial_{x}+S_{\chi}
   \partial_v - \left[ nz\left(1+\frac{vz}{x} \right) S_{\chi\chi}
   + \frac{nz}{x} S_{\chi} \right] \partial_{n} \,.    \end{equation}
Here the function $S$, defined by the form of the intensity
boundary distribution,
 \[
 S(\chi) = \alpha N(\chi) + \frac{\beta}{\chi\sqrt{N(\chi)}} \,
 \partial_{\chi} \left( \chi \partial_{\chi} \sqrt{N(\chi)}
  \right) \, ,
 \]
contains two small parameters, $\alpha$ and $\beta$, and, as in
the case $\beta=0$, there exist specific forms of boundary
distribution, $N$, for which the RGS operator (\ref{rgsym})
defines exact (not approximate) symmetry valid for arbitrary
values of $\alpha$ and $\beta$. \par
  Constructing a particular BVP solution (step \textbf{IV}) implies the
use of group invariants related to (\ref{rgsym}), and the
procedure is similar to that one for the parabolic beam. For the
Gaussian wave beam, $N=\exp (-x^2)$, the result is as follows:
\begin{equation}   \label{res-gs1}
 v(z,x)= \frac{x-\chi}{z}, \quad  n(z,x)= e^{-\mu^2} \frac{\chi}{x}
        \frac{\left( \beta - \alpha e^{-\chi^2}\right)}
    {\left( \beta - \alpha e^{- \mu^2} \right)} \,.    \end{equation}
Here $\chi$ and $\mu$ are expressed in terms of $t$ and $x$ by the
implicit relations
 $$
 \begin{array}{l}
 \displaystyle{ \beta \left( \mu^2 - \chi^2 \right)
 + \alpha \left( e^{- \mu^2} - e^{- \chi^2 }\right) =
 2z^2 \chi^2 \left( \beta - \alpha e^{-\chi^2} \right)^2 \,;} \\  \mbox{}\\
 \displaystyle{ x=\chi\left(1+ 2z^2\left(\beta -\alpha e^{-\chi^2} \right)
 \right) \,.}    \end{array} $$

The solution (\ref{res-gs1}) describes the self-focusing of the
cylindrical Gaussian beam that gives rise to the two-dimensional
singularity: both the beam intensity $n$ and derivatives $v_x,
n_x$ go to infinity at the point $z_{sing}^{Gauss} = 1/\sqrt{ 2
(\alpha - \beta)}$ provided that $\alpha > \beta$. A detailed
analysis of (\ref{res-gs1}) and more general solutions with a
parabolic form of an eikonal at $z=0$, $v(0,x)=-x/T$, is given in
\cite{K-TMF99,PRA}.

To illustrate the difference between the one-- and
two--dimensional solution singularities, in Figure \ref{optics} we
present a typical behavior of the wave beam intensity, defined by
Eq.(\ref{res2_gauss}) and (\ref{res-gs1}).
 \begin{figure}[hbtp]
 \centerline{\includegraphics[width=0.98\textwidth]{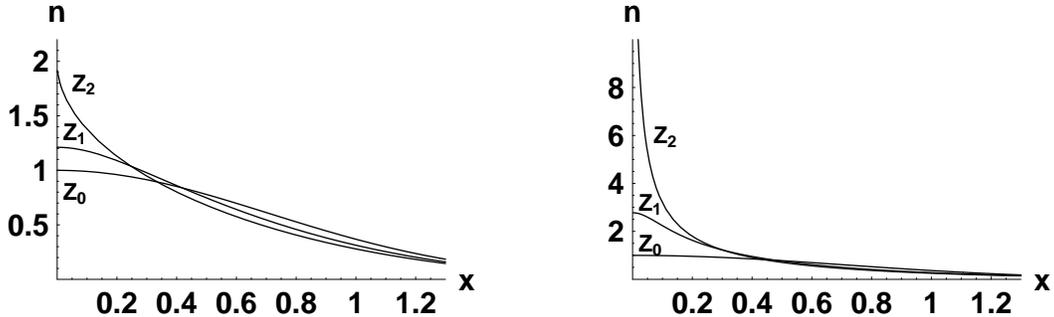}}
 \caption{Intensity $n$ versus transverse coordinate
$x$ for a plane (left panel) and cylindrical (right panel) beam
geometry for a few values of distance $z$ from the boundary
 $z_2>z_1>z_0=0$.} \label{optics}
 \end{figure}
The left panel corresponds to the plane beam geometry, $\nu=1$,
and without diffraction, $\beta=0$, while the right one is
concerned with a cylindrical wave beam, $\nu=2$, with both
nonlinearity and diffraction effects included. Diverse curves
describe beam intensity distribution upon coordinate $x$ at
different distances from the medium boundary, where we have the
collimated Gaussian beam, $N=\exp(-x^2)$. \par

It is clear that in the plane geometry the derivative of the beam
intensity with respect to $x$ turns to infinity at some singular
point, while the value of intensity on the axis remains finite. In
cylindrical case the solution singularity is two-dimensional: both
the beam intensity and its derivative with respect to transverse
coordinate turn to infinity simultaneously at point $z_{sing}$.
This last example demonstrates the possibility of the RGS approach
\textit{to analyse two-dimensional singularity}. In the practice
of RG application to critical phenomena, this correlates with the
case of ``two renormalization groups"\cite{cris98}.

\section{Overview}

To complete our review, we indicate  milestones in evolution of
the RG concept. Since its appearance in QFT,  RG served as a
powerful tool of analyzing diverse physical problems and improving
solution singularities disturbed by perturbation approximation.

The development of the RG concept can be divided into two stages. The
first one (since the mid-50s up to the mid-80s) is summarized in Figure 3.
 \begin{figure}[hbtp]
 \centerline{\includegraphics[width=0.8\textwidth]{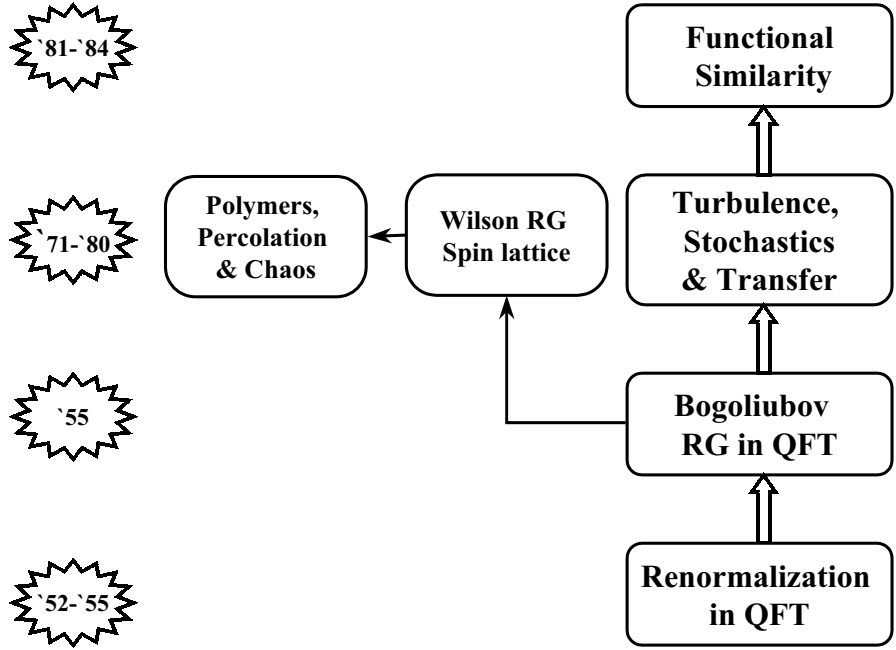}}
 \caption{Early development of concept: from Bogoliubov RG to Wilson RG and
 FS.} \label{fig3}
 \end{figure}
Besides early history (discovery of RG, formulation of the RG
method and application to UV and IR asymptotics), it comprises the
devising of the Kadanoff--Wilson RG in the 70s and following explosive
expansion into other fields of theoretical physics. \par

During this stage, the formulation of the RG method was based on
the unified scaling transformation of an independent variable
(and/or some parameters) accompanied by a more complicated
transformation of a solution characteristic $g_{\mu}=
\bar{g}(\mu,g)$ -- see Eqs.({\ref{rgt-m0}), (\ref{1-12}) and
(\ref{1-17}) in Section \ref{ss1.2}.

Here, the main role of RG=FS was the {\it \'a priori}
establishment of the fact that the solution under consideration
admits functional transformations that form a group.

Any particular implementation of the RG symmetry differs in the
form of the function(s) $\bar{g}(\mu, g)$ (or $\beta
(g)$) which, in an every partial case, is obtained from some
approximate solution.
\smallskip

The next stage, after the mid--80s, is depicted in the Figure 4. The
scheme describes the entire evolution of the Bogoliubov RG.
 \begin{figure}[hbtp] \label{fig4}
 \centerline{\includegraphics[width=0.8\textwidth]{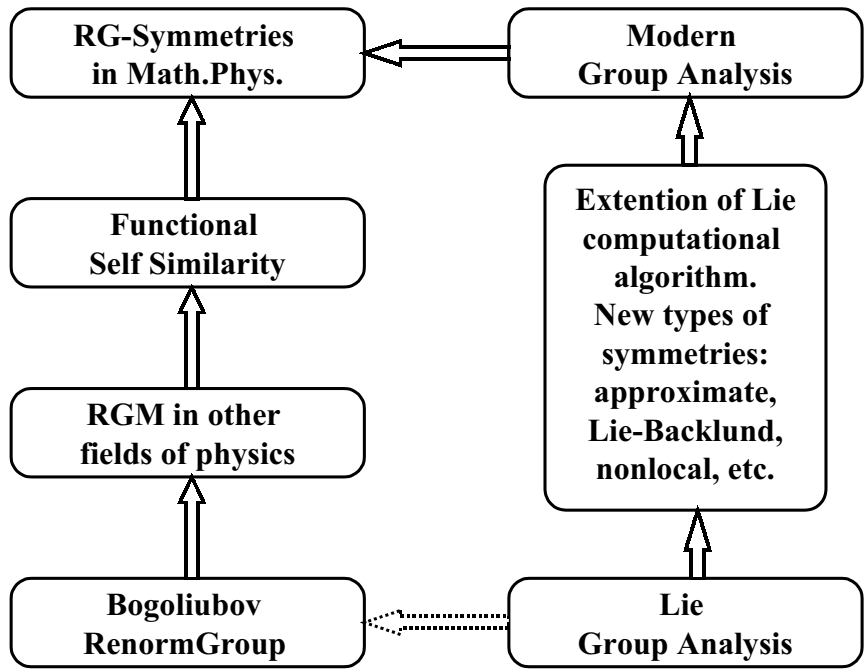}}
 \caption{Evolution of concept: from the Bogoliubov RG via FS to
RG-symmetries.} \end{figure} There were several important reasons
in further devising the RG concept in theoretical physics at this
period. On the one hand, it was due to the extension of the notion
of FS and RG symmetry that until then were based on one-parameter
Lie group of point transformations. Appending multi-dimensional
Lie point groups and Lie--B\"acklund groups to possible
realization of group symmetry enhanced the capability of RG
method. On the other hand, this additional possibility arose due
to the mathematical apparatus that was used in mathematical
physics to reveal RGS.  The advantage came from infinitesimal
transformations that enabled to describe RGS by an algebra of RG
generators.  However, in contrast to the situation typical of QFT
models with only one operator, in mathematical physics we have
finite or infinite-dimensional algebras. Both their dimension and
the method of construction depend upon a model employed and upon a
form of boundary conditions. \par

The use of infinitesimal approach results in constructing the
RG-type symmetry with the help of regular methods of group
analysis of DEs. Precisely, this regular algorithm naturally
includes the RG=FS invariance condition in the general scheme of
constructing and application of RGS generators (see also our
recent review \cite{KSh-TMF99}). Within the infinitesimal approach
this condition is formulated in terms of vanishing of canonical RG
operator coordinates, which is especially important for
Lie--B\"acklund RGS because finite transformations in this case
are expressed as formal series. In particular, this property
attribute a new feature to the RG analysis of a BVP solution with
singular behavior, making a singularity analysis more powerful.

At the same time, as the group analysis technique is still
developing -- here we mean both extension to new types of
symmetries and application to more complicated mathematical
models, e.g., including integro--differential equations -- we have
a clear perspective that the possibilities of a regular scheme
based upon the Bogoliubov renormalization group method are far
from being exhausted.

\medskip
{\large\textbf{ Acknowledgments} }
\smallskip

The authors are grateful to Professors Cris Stephens and Denjoe
O'Connor for invitation to participate in the Conference
``Renormalization Group 2000". They are indebted to these
gentlemen for useful discussions and comments. This work was
partially supported by grants of the Russian Foundation for
Basic Research (RFBR projects Nos 96-15-96030, 99-01-00232
and 99-01-00091) and by INTAS grant No 96-0842, as well as by the
Organizing Committee of the above-mentioned meeting.

\end{document}